\documentclass{aa}  

\usepackage{graphicx}
\usepackage{txfonts}
\usepackage{natbib}
\bibpunct{(}{)}{;}{a}{}{,}
\usepackage{subfig}

\begin{document}

\title{Investigation into the reflection properties of the neutron star low-mass X-ray binary 4U 1636$-$53}

\author{Ming\ Lyu  \inst{1,2}
  \and 
Guobao \ Zhang \inst{3,4}
  \and 
H. G. Wang \inst{5}
  \and
Federico Garc\'{\i}a \inst{6}
}
\institute{Department of Physics, Xiangtan University, Xiangtan, Hunan 411105, China
 \and
 Key Laboratory of Stars and Interstellar Medium, Xiangtan University, Xiangtan, Hunan 411105, China\\
 \email{lvming@xtu.edu.cn}
 \and
 Yunnan Observatories, Chinese Academy of Sciences (CAS), Kunming 650216, P.R. China
 \and
 Key Laboratory for the Structure and Evolution of Celestial Objects, CAS, Kunming 650216, P.R. China
  \and
 Department of Astronomy, School of Physics and Materials Science, Guangzhou University, Guangzhou 510006, China 
  \and
Instituto Argentino de Radioastronom\'ia (CCT La Plata, CONICET; CICPBA; UNLP), C.C.5, (1894) Villa Elisa, Buenos Aires, Argentina
 }

\abstract
%context heading (optional)
{We present  the spectroscopy of the neutron star low-mass X-ray binary 4U 1636--53 using six simultaneous XMM–Newton and Rossi X-ray Timing Explorer observations. We applied different self-consistent reflection models to explore the features when the disk is illuminated by either the corona or the neutron star surface. We found that the spectra could be well fitted by these two types of models, with the derived emissivity index below a typical value of 3. The relative low emissivity can be explained if the neutron star and the corona, working together as an extended illuminator, simultaneously illuminate and ionize the disk. Additionally, the derived ionization parameter in the lamppost geometry is larger than the theoretical prediction. This inconsistency likely suggests that the corona does not emit isotropically in a realistic context. Furthermore, we also found that there is a possible trend between the height of the corona and the normalization of the disk emission. This could be understood either as a variation in the reflected radiation pressure or in the context of a jet base. Finally, we found that the disk is less ionized if it is illuminated by the neutron star, indicating that the illuminating source has significant influence on the physical properties of the disk.
}

\titlerunning{Investigation into 4U 1636$-$53}
\authorrunning{Lyu et al.}
\maketitle
%________________________________________________________________

\section{Introduction}  

A neutron star low-mass X-ray binary system (NS LMXB) is composed of a neutron star and a low-mass (M$\le$1 M$_\odot$) companion star. The matter is accreted from the companion to the central neutron star surface through the accretion disk. Historically, NS LMXBs were classified as   Atoll types or   Z types, according to their color patch in an X-ray color-color diagram \citep{hasinger89}. The Atoll source has a lower luminosity  compared with the Z source  \citep[e.g.,][]{ford2000,done07,homan07}. In the color-color diagram, the Atoll source displays three branches: the island, the lower banana, and the upper banana. These three branches correspond to the hard, transitional, and soft spectral states, respectively.  It is proposed that the mass accretion rate increases when the source evolves from the hard spectral state to the soft spectral state.

In an NS LMXB system, the thermal radiation comes from the neutron star surface and the inner part of the accretion disk, while the  nonthermal emission is generated in the  corona where the thermal photons are inverse-Compton scattered by the hot electrons. Furthermore, photons emitted from the corona and the neutron star surface illuminate the top layer of the accretion disk, and hence produce the reflection component of the disk. In this process, some photons are absorbed and reprocessed by the disk and, as a result, some absorption and emission line features appear in the reflection spectrum. Among these features, the most prominent one is the fluorescent iron emission line in 6.4 - 6.97 keV, due to its large cosmic abundance and the high fluorescent yield. Investigation in the reflection features provides a key tool to study the geometry and the physical state of the accreting system \citep[e.g.,][]{Bhattacharyya07,Cackett08,ludlam17,ludlam22}.

The binary 4U 1636$-$53 is an atoll NS LMXB with a companion star of mass $\sim$ 0.4 M$_\odot$ \citep{giles02} in a 3.8 h orbit \citep{van90} at a distance of 6 kpc \citep{galloway06}. 4U 1636--53 harbors a millisecond pulsar with a spin frequency of 581 Hz \citep{zhang97,strohmayer02}. The source has exhibited type I nuclear X-ray bursts, super-bursts \citep{strohmayer02}, quasi-periodic oscillations (QPOs) \citep{zhang96} and time lags \citep{kaaret99}. Additionally, 4U 1636--53 shows the full range of the spectral states \citep{belloni07,diego08}, with a regular state transition cycle of  $\sim$ 40 d \citep{shih05,belloni07}.

The reflection feature of 4U 1636$-$53 has been extensively investigated in   previous works. \citet{pandel08} studied the broad and asymmetric emission lines and proposed that the profiles are due to the blending of two Fe-K lines at different energies. \citet{cackett10} found that the lines could be well fitted by a reflection model assuming that the blackbody component from the boundary layer of the neutron star illuminates the accretion disk. Later, \citet{sanna13} found that the primary source that illuminates the disk is the neutron star and its boundary layer in four observations, while in the remaining two observations the dominant illuminating source is the corona. \citet{lyu14} found that the flux and equivalent width of the iron line first increase and then decrease as the flux of the Comptonized component increases, which could be due to the light bending effect or variation in  the disk ionization state. \citet{ludlam17} studied reflection of 4U 1636--53 with a NuSTAR observation and constrained the inner radius of the disk to be 1.08$\pm$0.06 $R_{ISCO}$ in the hard spectra state. \citet{wyn17} found that the inclination angle of 4U 1636--53 is higher than 80 degrees in the reflection model {\sc relxill}, and that the height of the corona in the lamppost geometry is very small, $\sim$2.5 $R_{g}$. Recently, \citet{mondal21} found a clear signature of disk reflection, a broad Fe-K emission line, and a reflection hump in one NuSTAR observation of 4U 1636--53 in the hard state. Furthermore, they derived an inner radius of the disk: $\sim$ 3.2-4.7 $R_{ISCO}$.

For this work we investigated the reflection properties of 4U 1636--53 when the source was in different spectral states. We applied full reflection models to explore spectral features when the disk is illuminated by either the neutron star surface (plus its boundary layer) or the corona. Furthermore, we studied the ionization state of the disk by comparing the results from spectroscopy and theory. The paper is organized as follows. We show the observations and the data reduction process in Section 2. We describe the details of the spectral analysis in Section 3, and show the corresponding results in Section 4. In Section 5, we briefly discuss the derived results.

\section{Observations and data reduction}

For this work we analyzed six XMM-{\it Newton} observations plus Rossi X-ray Timing Explorer ({\it RXTE}) observations of 4U 1636--53. The XMM-{\it Newton} observations were operated with EPIC-PN \citep{struder01} in timing mode, with the readout speed increased to suppress the pileup effect. The {\it RXTE} observations were taken simultaneously with the XMM-{\it Newton} observations (for  details of the XMM-{\it Newton/RXTE} observations, see  Table \ref{obs}).

\subsection{XMM-{\em Newton} data reduction}
We used the Science Analysis System (SAS) version 19.1.0 for the data reduction of the XMM-{\it Newton} observations. We first applied the tool {\tt epproc} to reprocess the raw data and generate calibrated EPIC event lists, and then ran the tool {\tt barycen} to convert the arrival time of photons to the barycenter  of the Solar System. We selected only single and double events (PATTERN$<=$4), and excluded all events at the edge of CCD and close to a bad pixel (FLAG==0) for the spectra extraction. All X-ray bursts were removed before the extraction. To test the possible pileup effect, we ran the tool {\tt epatplot} and found that there is moderate pile-up in the XMM-{\it Newton} observations. We then selected a 41-column wide region centered on the source position, with the central 3 columns excluded for   observation X3 and X6. For the rest of the  observations, we excluded the central  column for the correction of the pileup effect. Additionally, in the PN timing mode, the whole CCD is full of the source photons \citep{ng10,hiemstra11} due to the fact that its point spread function (PSF) extends farther  than the boundaries. We then selected an observation of black hole binary GX 339--4 (ObsID 0085680601), in timing mode when the source was close to quiescence as the blank field for the background extraction. We ran the command {\tt rmfgen} and {\tt arfgen} to produce the response matrices and ancillary response files.  Finally, we re-binned all the source spectra by a factor of 3 to ensure that there was a minimum of 25 counts per bin with the tool {\tt specgroup}.

\subsection{{\em RXTE} data reduction}
For the {\it RXTE} data reduction, we used the {\sc heasoft} package version 6.28 following the steps in the RXTE cook book.\footnote{http://heasarc.gsfc.nasa.gov/docs/xte/recipes/cook\_book.html} We applied the tool {\tt saextrct} to reduce the Proportional Counter Array \citep[PCA,][]{jahoda06} data in the Standard-2 format from the best-calibrated PCU2 detector alone. We excluded the bursts and applied the dead-time correction during the extraction. The background spectra were extracted with the command {\tt pcabackest}, and the response files were generated with the tool {\tt pcarsp}. For the High Energy X-ray Timing Experiment \citep[HEXTE,][]{roth98} data, we used the cluster-B events only, with the detector 2 data excluded since it lost the ability to measure spectral information after 1996 March. We used the command {\tt saextrct} to extract the spectra and applied the tool {\tt hxtdead} for the dead-time correction. The response files were generated with the tool {\tt hxtrsp}. Finally, we used the tool {\tt grppha} to group all the PCA and HEXTE spectra so that there were at least 30 counts in each energy bin.

\begin{table*}
\centering
\tiny
\caption{XMM-{\em Newton/RXTE} observations of 4U 1636--53 in this work.}
\begin{tabular}{cccc}
\hline
Observation   &   Telescope      &  ObsID   & Start time      \\
\hline
X1 ($S_{a}$=1.33)    &  XMM-{\it Newton}  & 0303250201         & 2005-08-29 18:25  \\
         &   {\it RXTE}             & 91027-01-01-000  &2005-08-29 16:35  \\
\\
X2 ($S_{a}$=2.01)    &  XMM-{\it Newton}  & 0500350301          &2007-09-28 15:42  \\
         &   {\it RXTE}             &93091-01-01-000   &2007-09-28  14:47  \\
\\
X3 ($S_{a}$=2.08)     & XMM-{\it Newton}  & 0500350401          &2008-02-27 04:14  \\
         &   {\it RXTE}             &93091-01-02-000   &2008-02-27 03:46  \\
\\
X4 ($S_{a}$=2.17)     & XMM-{\it Newton}  & 0606070201          & 2009-03-25 23:01  \\
         &  {\it RXTE}             & 94310-01-02-03     & 2009-03-25 23:00 \\
         &                         & 94310-01-02-04    & 2009-03-26 00:39   \\
         &                         & 94310-01-02-05    & 2009-03-26 02:17   \\
         &                         & 94310-01-02-02    &  2009-03-26 03:54  \\
 \\
X5 ($S_{a}$=2.09)     & XMM-{\it Newton}  & 0606070301         & 2009-09-05 01:57  \\
         &  {\it RXTE}             & 94310-01-03-000  & 2009-09-05 01:17 \\
         &                         &  94310-01-03-00   & 2009-09-05 08:20 \\
\\
X6 ($S_{a}$=1.36)     & XMM–{\it Newton}  & 0606070401         & 2009-09-11 08:47 \\
         &  {\it RXTE}              & 94310-01-04-00    & 2009-09-11 08:42  \\

\hline          
\end{tabular}   
\medskip        
\\                  
\label{obs}     
\end{table*}

\begin{figure*}
\centering
\includegraphics[width=0.4\textwidth]{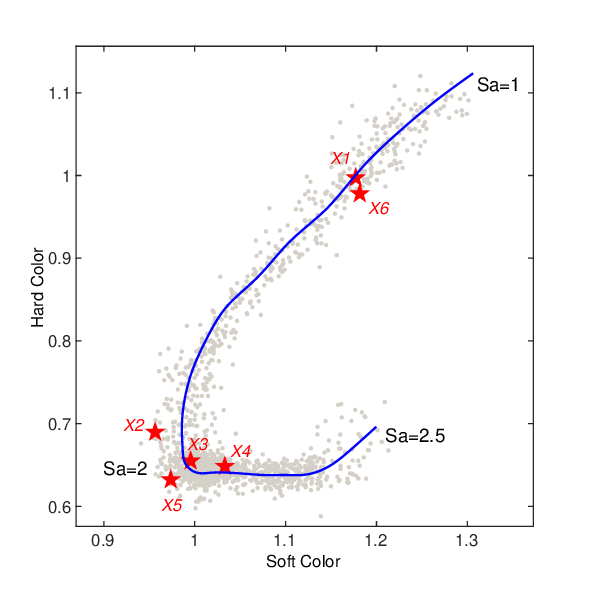}
\caption{Color–color diagram of 4U 1636--53. Each gray point represents a single {\it RXTE} observation. The red stars indicate the six observations in this work. The position of the source in the diagram is parameterized by the length of the blue solid curve S$_{a}$.}
\label{ccd}
\end{figure*}

Furthermore, we made a color-color diagram to trace the spectral state of the six observations. We used 16 s time-resolution PCA Standard-2 data for the calculation of the two colors, with the hard and soft colors defined as the 9.7–16.0/6.0–9.7 keV and the 3.5–6.0/2.0–3.5 keV count rate ratio, respectively. The colors were then normalized to those of the Crab nebula in observations taken close in time to those of 4U 1636--53 \citep[see][for more details]{zhang09}. We selected the parameter S$_{a}$ to trace the location of the source in the color–color diagram \citep[see, e.g.,][]{mendez99,zhang11}. During the calculation, we used the same definition of S$_{a}$ as in \citet{zhang11}: S$_{a}$=1 corresponds to the top right vertex and S$_{a}$=2 to the bottom left vertex of the color–color diagram. In Figure \ref{ccd} we show the distribution of the six observations in the color-color diagram of 4U 1636--53. It is clear that observations X1 and X6 are located in the hard branch, while observations X2-X5 are in the soft branch.

\section{Spectral analysis}
 In this work we used the XSPEC version 12.12.0 \citep{arnaud96} to fit the PN, PCA, and HEXTE spectra together for the hard observations X1 and X6 in the $0.8-80.0$ keV energy range (PN:$0.8-10.0$ keV; PCA:$10.0-25.0$ keV; HEXTE:$25.0-80.0$ keV). For   observations X2--X4 in the soft state we only selected the PN and PCA spectra since there are very few source photons in the HEXTE energy range. We excluded the $1.5-2.5$ keV range for the PN spectra to avoid possible calibration problems in this range. A systematic error of 0.3\% was used in the fits to account for the possible cross-calibration effects on the spectra shape. We selected the component {\sc tbabs} to describe the absorption due to the interstellar medium along the line of sight, with the solar abundance of \citet{wilms00} and the photon-ionization cross section table of \citet{verner96}. In the fitting process we applied a multiplicative factor, which is fixed at 1 for the PN spectra and is free for others.   
 
 We first tried a simple model ({\sc bbody+diskbb+cutoffpl}) to estimate the shape of the spectra. The {\sc bbody} component   describes the thermal emission from the neutron-star surface, and the {\sc diskbb} component \citep{mitsuda84,maki86}   fits the thermal radiation from the accretion disk. We used the {\sc cutoffpl} model to account for the nonthermal emission from the corona. Additionally, a {\sc gauss} model was selected to fit the possible iron line emission in the energy range 6.4-6.97 keV. We found that the iron line is significant in all observations, consistent with the finding in \citet{sanna13}. We then used two types of models to investigate the  reflection off the accretion disk, as described below.

 \subsection{Corona illumination model}
 In LMXBs the accretion disk could be illuminated and ionized by photons from the corona. To explore this scenario, we  applied a self-consistent model ({\sc bbody+diskbb+relxillCp}) where   {\sc relxillCp} \citep{garcia14,dauser16} calculates the reflection from the accretion disk illuminated by a Comptonization power-law continuum. The {\sc relxillCp} model combines the  {\sc xillver} \citep{garcia10,garcia13} reflection code and the {\sc relline} \citep{dauser10,dauser13} ray tracing model, and is able to calculate the reflection at each emission angle. The parameters in the {\sc relxillCp} model are the emissivity index of the inner and outer disk, $\beta_{in}$ and $\beta_{out}$; the breaking radius, $R_{br}$, where the emissivity changes; the spin parameter, $a$; the redshift, $z$; the inclination angle; the inner and outer radius of the disk, $R_{in}$ and $R_{out}$; the photon index, $\Gamma$; the ionization parameter, $\xi$; the iron abundance; the cut-off energy of the power law, $E_{cut}$; the reflection fraction, $R_{refl}$; the density of the accretion disk; and the normalization. In the fits the iron abundance was fixed at the solar abundance and the redshift $z$ was set to 0. We fixed the spin parameter $a$ at 0.27 calculated as $a$ = 0.47/$P_{ms}$ \citep{braje00}, where the spin period of the neutron star in milliseconds is $P_{ms}$=1000/581 ms. We set the breaking radius $R_{br}$ to be the same as the outer radius $R_{out}$=400 $R_{g}$, and then linked the emissivity $\beta_{in}$ and $\beta_{out}$. The inner radius $R_{in}$ is set to be larger than 5.12 $R_{g}$ and the density of the disk is fixed at a value of 10$^{15}$ cm$^{-3}$ in the fits. The inclination angle in the fits always goes close to $\sim$90 degrees, which is unphysical since there is no eclipse observed in the light curve of this source. Previous optical measurements in \citet{casares06} indicate that the inclination angle in this source is $\sim$ 60--71 degrees. We thus fixed the inclination angle in the fits at 70 degrees in this work. 
  
We further applied other reflection models {\sc relxilllp} and {\sc relxilllpCp} to study the geometry of the source. In the {\sc relxilllp} model it is assumed that the corona is a point source located at a height above the central compact object. Most of the parameters in the {\sc relxilllp} model are the same as in {\sc relxill}, but instead of the emissivity index, {\sc relxilllp} has two new parameters, $h$ and $\beta$, which is the height of the corona and the corona velocity. We fixed the $\beta$=0 since it is not well constrained in the fits. Moreover, we set the parameter $fixReflFrac$ to its default value 0, and left the other parameters the same as in {\sc relxill}. In the {\sc relxilllpCp} model, the accretion disk is illuminated by a thermal Comptonization spectrum instead of a power-law spectrum in the {\sc relxilllp} model. We selected a constant ionization of the disk by setting the parameter $iongrad\_index$ to zero. The other parameters are the same as in the  {\sc relxilllpCp} model.

\subsection{Neutron star illumination model}

When the source is in the soft state, the accretion disk likely comes close to the neutron star, so the radiation from the neutron star  surface could illuminate the accretion disk and hence generate the reflection off the disk. Accordingly, we fit the reflection in the four observations (X2-X5) in the soft state with the model {\sc bbrefl} \citep{ballan04}, which calculates the reflection spectrum from a constant density disk  illuminated by a {\sc bbody} component. The parameters in {\sc bbrefl} are the ionization parameter of the disk, $\xi$; the temperature of the incident blackbody photons, $kT$; the redshift, $z$; and the normalization. In the fits we convolved the {\sc bbrefl} with the model {\sc relconv} \citep{dauser10} to account for the relativistic effects around the compact object. The model is parameterized by  the spin, $a$; the inclination angle; the inner and outer radius, $R_{in}$ and $R_{out}$; the breaking radius, $R_{br}$; and the two emissivities, $\beta_{in}$ and $\beta_{out}$. We set the temperature $kT$ in {\sc bbrefl} to be the same as the $kT_{\rm BB}$ in the  {\sc bbody} component. The settings of the parameters $a$, $z$, $R_{in}$, $R_{br}$, $R_{out}$, $\beta_{in}$, and $\beta_{out}$ are the same as  in the  {\sc relxill} model. In the fits we found that the value of  $R_{in}$ could not be constrained, so we fixed it at one marginally stable radius in the model.

Recently, a new model, {\sc relxillNS}, was developed to explain the reflection from the accretion disk illuminated by a blackbody component from the neutron star surface. We thus also applied the model {\sc diskbb+cutoffpl+relxillNS} for the soft observations in this work. The parameters in the  {\sc relxillNS} model are the same as   in the  {\sc relxillCp} model, except that there is a new parameter, $kT_{\rm BB}$, which  replaces the parameters $E_{cut}$ and $\Gamma$ in the {\sc relxillCp} model. We set the reflection fraction parameter to be positive so that the {\sc relxillNS} component describes the neutron star radiation and the reflection together in the fits. We set the other parameters in the fits to be the same as in the  {\sc relxillCp} model.\\

To sum up, we applied the models {\sc bbody+diskbb+relxillCp}, {\sc bbody+diskbb+relxilllp}, and {\sc bbody+diskbb+relxilllpCp} to fit the spectra when the disk is illuminated by the corona. Instead, we used the models {\sc bbody+diskbb+cutoffpl+relconv$\times$BBrefl} and {\sc diskbb+cutoffpl+relxillNS} to explore the physics when the disk is illuminated by a neutron star surface.

\section{Results}

\begin{figure*}
\center
\includegraphics[width=0.4\textwidth]{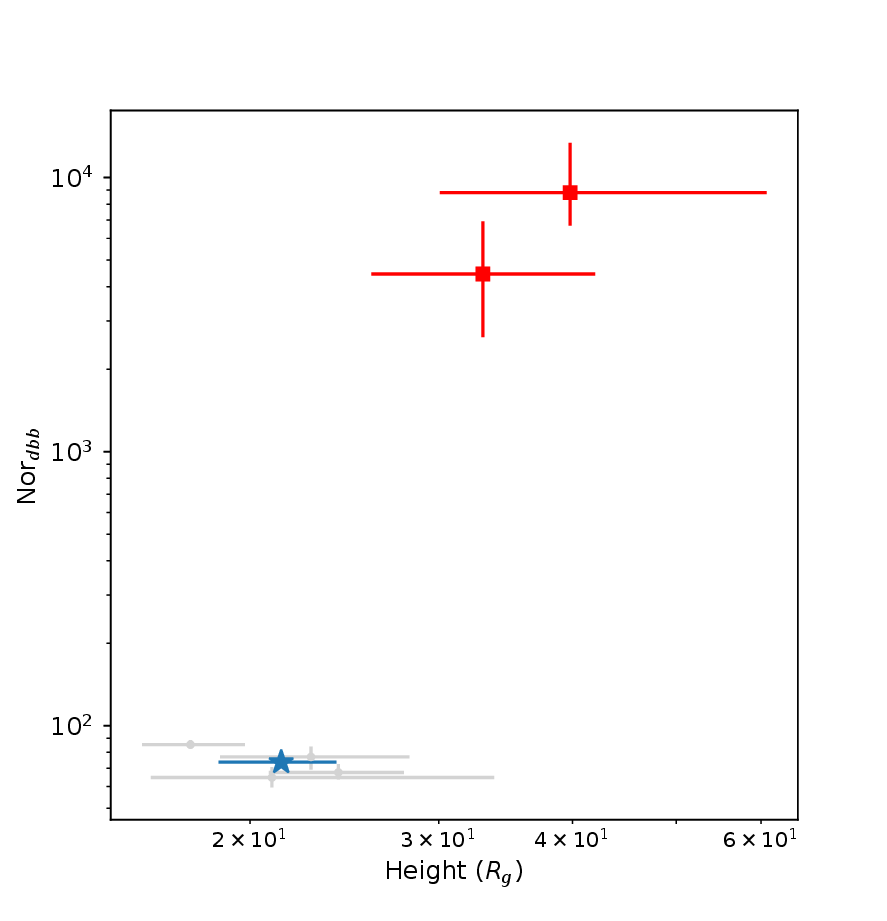}
\caption{Normalization of  {\sc diskbb} component vs.  height of the corona for the observations of 4U 1636--53 in this work. Here are shown 1$\sigma$ error bars, and  the hard observations (X1 and X6) and soft observations (X2--X5) are indicated with red squares and gray circles, respectively. The average of the soft observations is shown as a blue star for clarity.}
\label{h_N}
\end{figure*}
 
In Table \ref{relxill} we show the fitting results with the  {\sc relxillCp} model. The blackbody temperature, $kT_{\rm BB}$, is around 2.1 keV for   X2--X5, while it is 1.85$\pm 0.11$ keV and 2.39$\pm 0.08$ keV for   X1 and X6. The accretion disk temperature, $kT_{\rm disk}$, is in the range of $\sim$ 0.9--1.0 keV when the source is in the soft state, while it is around 0.22 keV in the hard state. For the normalization of the {\sc diskbb}, it is less than 90 in the observation X2-X5, while it is greater than $\sim$ 7000 and $\sim$ 2500 in   X1 and X6, suggesting that the accretion disk is far away from the neutron star when the source is in the hard state. The power-law index, $\Gamma$, is around 1.8 in the two hard observations, and it is a little bigger, consistent with a value of $\sim$ 2.2 in the soft state. The electron temperature in observations X1 and X6 is 18$\pm$2 keV and 42$_{-14}^{+29}$ keV, respectively, while it is much lower, $\sim$ 10 keV, in the four soft observations. For the ionization of the accretion disk, log($\xi$), it is distributes over a range from 3.27 to 3.87 in these six observations, corresponding to an ionization parameter from 1860 erg$\cdot$cm/s to 7410 erg$\cdot$cm/s. The emissivity indices in observations X1 and X6 are around 1.7, and close to $\sim$ 2.1 in the soft observations, except that it is 2.7$\pm$0.5 in   X3.

In Table \ref{relxilllp} we show the fitting result with the  {\sc relxilllp} model. The blackbody temperature and the accretion disk temperature are close to the values in the  {\sc relxillCp} model. The normalization of the {\sc diskbb} component in  X1 and X6 is greater than 1500, significantly larger than the range of 55--90 in the four soft observations. The power-law index, $\Gamma$, is around 1.7 in the two hard observations, and its value is in the range  1.8--2.1 in the soft observations. The value of the ionization parameter, log($\xi$), is in the range from 3.33 to 3.76, corresponding to the parameter $\xi$ in the range  2130--5750 erg$\cdot$cm/s. The height of the corona is in the range of 14--41 $R_{g}$ in the soft state, while it is 40$_{-15}^{+55}$ $R_{g}$ and 33$_{-11}^{+18}$ $R_{g}$ in   X1 and X6, respectively. Furthermore, as shown in Figure \ref{h_N}, we found that there is likely a trend between the distribution of the height of the corona and the normalization of the {\sc diskbb} component, which is proportional to the square of the inner radius of the accretion disk. In the plot we see that the four soft observations cluster in the bottom left corner, while the two hard observations are in the top right-hand corner. Since the error bars are relatively large, we further averaged the four soft observations in the plot for clarity.

\begin{figure*}
\center
\includegraphics[width=0.95\textwidth]{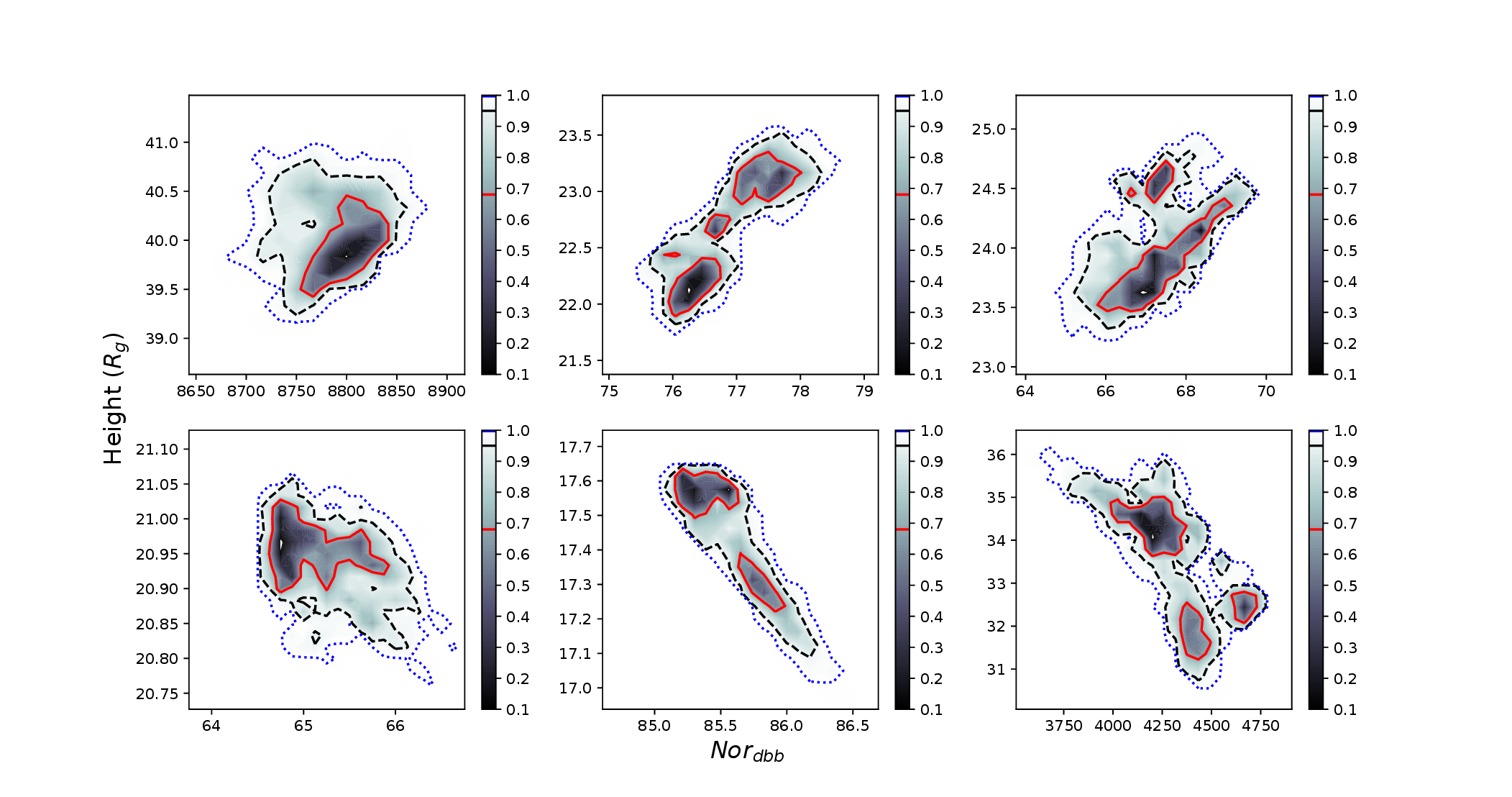}
\caption{Normalization of   {\sc diskbb} component vs.   height of the corona calculated from    MCMC techniques for the six observations of 4U 1636--53. The Goodman-Weare algorithm was used with eight walkers and a total length of 10000 steady-state samples for the simulation of each observation, with the initial 2000 chains removed to avoid an unstable state. For clarity, the 68\%, 95\%, and 99.7\% confidence levels are indicated as the red solid line, black dashed line, and blue dotted line, respectively.}
\label{mcmc}
\end{figure*}

\begin{figure*}
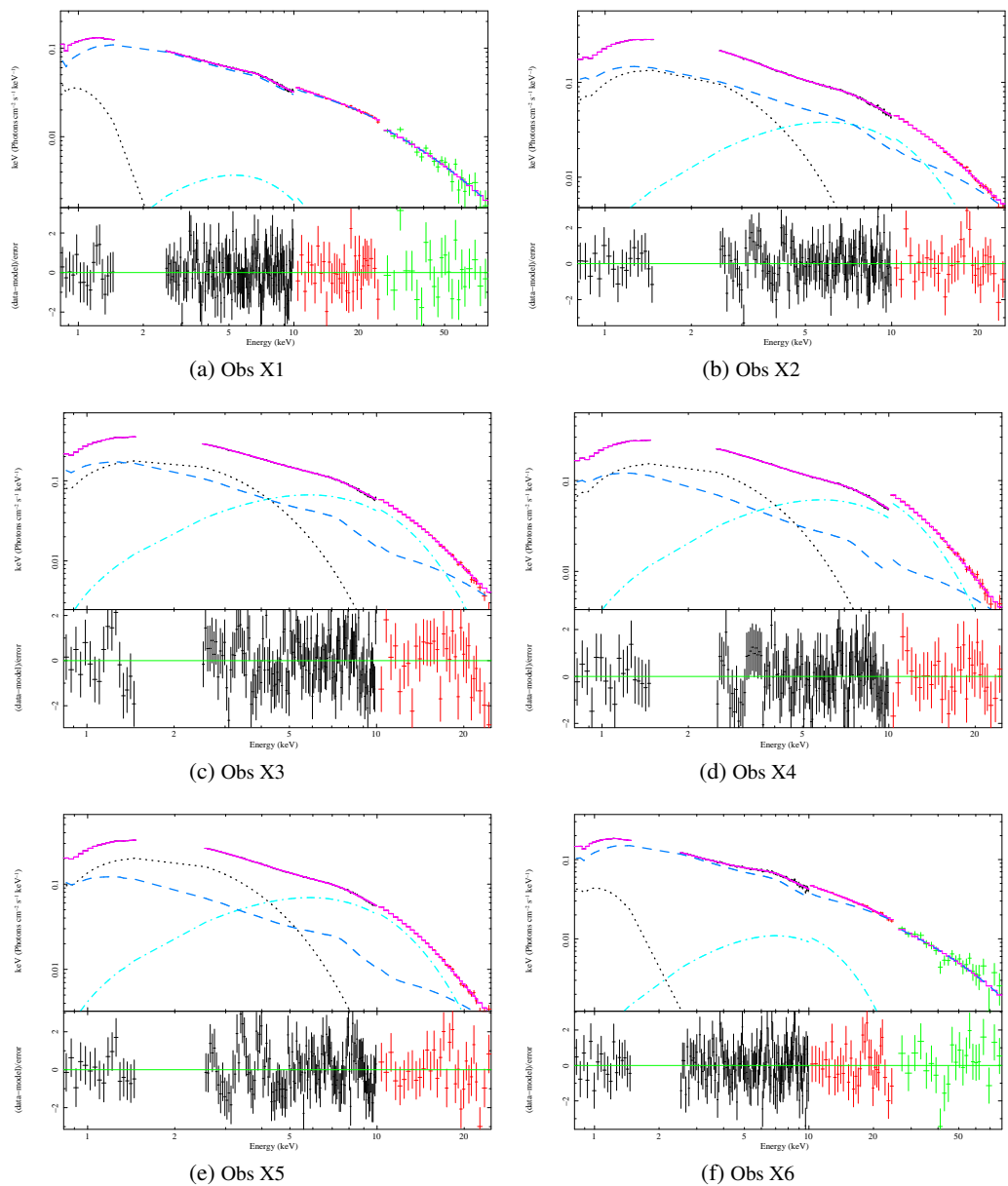

\center

\subfloat[\tiny Obs X1]{
\includegraphics[width=0.25\textwidth,angle=-90]{lp_o1.eps} }
\subfloat[\tiny Obs X2]{
\includegraphics[width=0.25\textwidth,angle=-90]{lp_o2.eps}}
\\
\subfloat[\tiny Obs X3]{
\includegraphics[width=0.25\textwidth,angle=-90]{lp_o3.eps}}
\subfloat[\tiny Obs X4]{
\includegraphics[width=0.25\textwidth,angle=-90]{lp_o4.eps}}
\\
\subfloat[\tiny Obs X5]{
\includegraphics[width=0.25\textwidth,angle=-90]{lp_o5.eps}}
\subfloat[\tiny Obs X6]{
\includegraphics[width=0.25\textwidth,angle=-90]{lp_o6.eps}}
\caption{Fitting results with the reflection model {\sc bbody+diskbb+relxilllp} for the XMM–Newton/RXTE observations of 4U 1636--53. The fitted spectra and the individual model components (main panel), and the residuals in terms of sigmas (subpanel) for the six observations are shown. The components {\sc bbody}, {\sc diskbb}, {\sc relxilllp}, and the sum of all these model components are plotted with a cyan dash-dotted line, black dotted line, blue dashed line, and purple line, respectively.}
\label{fit1}
\end{figure*}

 We also derived the distribution of the normalization of the  {\sc diskbb} component and the height of the corona in each observation from Markov Chain Monte Carlo (MCMC) in Xspec. As shown in Figure \ref{mcmc}, the range of the parameter $N_{dbb}$ and $h$ is $\sim$ 3600-8900 and $\sim$ 30-41 $R_{g}$ for the observations (X1 and X6) in the hard state, while the range is $\sim$ 64-87 and $\sim$ 17-25 $R_{g}$ for the observations (X2-X5) in the soft state. The results obtained from MCMC are consistent with the scenario indicated by Figure \ref{h_N} that both the accretion disk and the corona are at positions farther away from the neutron star when the source was in the hard spectra state. In Fig. \ref{fit1}, we show the corresponding spectra, individual components, and residuals of the fit with the  {\sc relxilllp} model.

We show the fitting result with   {\sc relxilllpCp} in Table \ref{relxilllpcp}. The power-law index, $\Gamma$, is around $\sim$ 1.85 in the hard state and around $\sim$ 2.2 in the soft observations. The values of the ionization parameter $\xi$ are distributed over a range of 2140--7760, consistent with the range derived in the  {\sc relxilllp model}. The value of the fitted corona height tends to be lower in the soft state than in the hard state, with its value pegged at its upper limit of 100 in  X1. The electron temperature, $kT_{e}$, is around 10 keV for the soft observations, while it is 24$_{-5}^{+10}$ keV and 41$_{-14}^{+46}$ keV in  X1 and X6.

 We show the fitting results using the neutron star illuminating model {\sc bbrefl} for the four soft observations in Table \ref{bbrefl}, with the corresponding spectra, individual components, and residuals in Fig \ref{fit2}. For these four observations in the soft state, the temperature $kT_{\rm BB}$ is $\sim$ 2.1 keV, and the disk temperature is $\sim$ 0.7-1.0 keV. The normalization of the {\sc diskbb} component is in the range from 47 to 154, consistent with the values in the corona illumination case, while the ionization parameter values, log($\xi$)$ \sim$ 3.2, in these four observations are smaller than the corresponding values derived in the corona illumination models. The emissivity of the disk is $\sim$ 2.3, close to the values derived in the model {\sc relxillCp}. 

\begin{figure*}
\center

\subfloat[\tiny Obs X2]{
\includegraphics[width=0.25\textwidth,angle=-90]{fitNS_o2.eps}}
\subfloat[\tiny Obs X3]{
\includegraphics[width=0.25\textwidth,angle=-90]{fitNS_o3.eps}}
\\
\subfloat[\tiny Obs X4]{
\includegraphics[width=0.25\textwidth,angle=-90]{fitNS_o4.eps}}
\subfloat[\tiny Obs X5]{
\includegraphics[width=0.25\textwidth,angle=-90]{fitNS_o5.eps}}

\caption{Fitting results with the reflection model {\sc bbody+diskbb+cutoffpl+relconv$\times$bbrefl} for the XMM–{\it Newton/RXTE} observations of 4U 1636--53 in soft spectral states. The fitted spectra and the individual model components (main panel), and the residuals in terms of sigma (subpanel) are shown. The components {\sc bbody}, {\sc diskbb}, {\sc cutoffpl}, {\sc bbref,l} and the sum of all these model components are plotted with a cyan dash-dotted line, black dotted line, red dash-triple-dotted line, blue dashed line, and purple line, respectively.}
\label{fit2}
\end{figure*}

\begin{figure*}
\center
\includegraphics[width=0.5\textwidth]{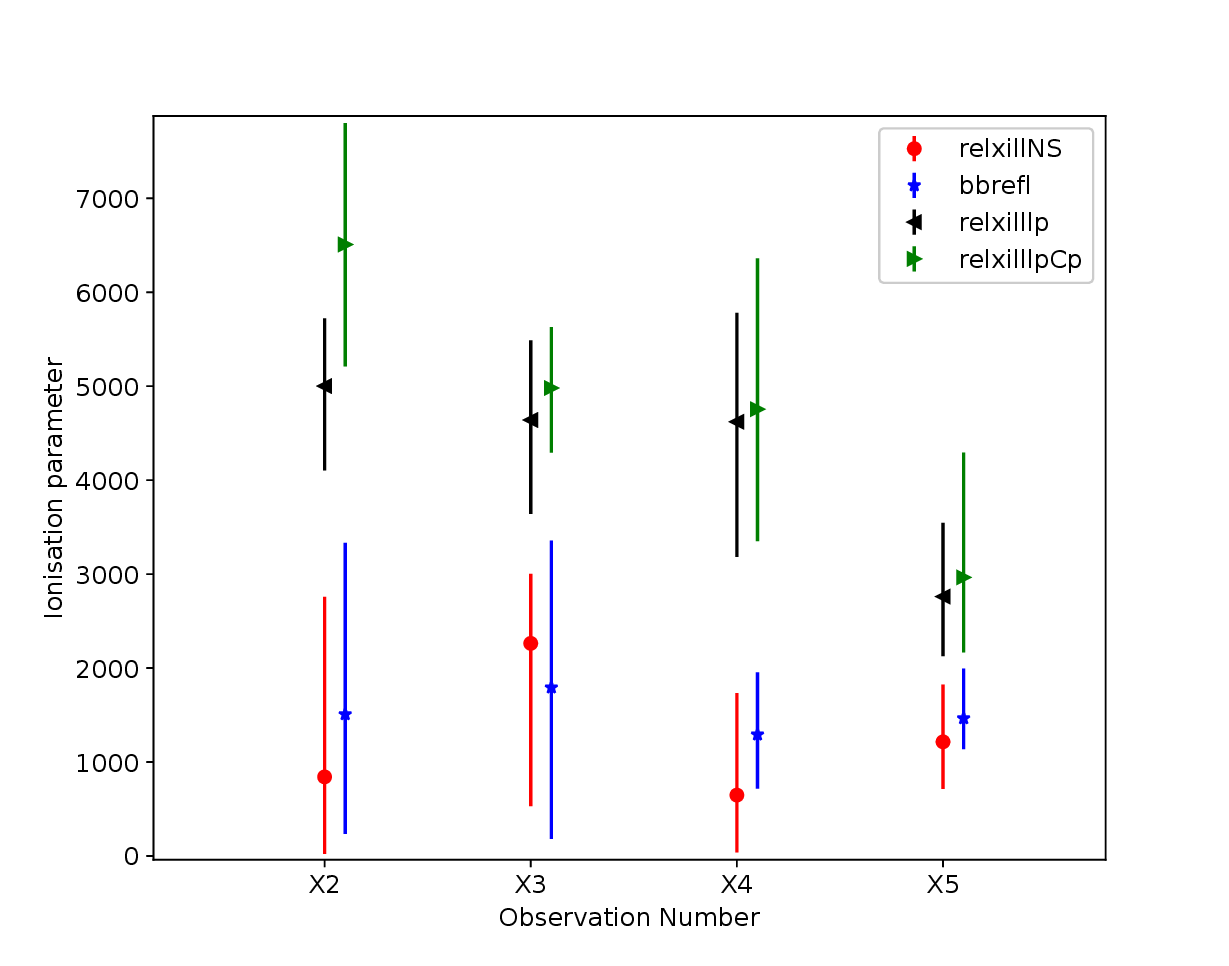}
\caption{Comparison of  ionization parameter derived from the neutron star illumination models ({\sc relxillNS}, {\sc bbrefl}) and the corona illumination models ({\sc relxilllp}, {\sc relxilllpCp}) for the four observations (X2-X5) in the soft state. Uncertainties are given at a 90\%\ confidence level unless otherwise indicated.}
\label{ioniza}
\end{figure*}

In Table \ref{relxillns} we show the fitting results with the  {\sc relxillNS} model. The fitted parameters in the {\sc relxillNS} model are very close to those obtained in the  {\sc bbrefl} model. In Figure \ref{ioniza} we compare the ionization parameter derived with the two types of illuminating models for the soft observations in this work. The ionization parameter deduced in the  {\sc relxilllp} model is consistent with that in  {\sc relxilllpCp}, and the ionization parameter in the model {\sc bbrefl} agrees closely with that in the  {\sc relxillNS} model. However, as shown in the figure, the ionization parameters from the two neutron star illuminating models are clearly lower than those derived in the two corona illuminating models.

\begin{table*}
\centering
\caption{Best-fitting results for the fit to the X-ray spectra of 4U 1636--53 with the reflection model {\sc bbdoy+diskbb+relxillCp}.}
\resizebox{\textwidth}{42mm}{
\begin{tabular}{cccccccc}
\hline
\hline
Model Comp   &   Parameter     & X1       & X2       & X3       & X4       & X5        & X6    \\
\hline

{\sc Tbabs}    &$N_{\rm H}$ (10$^{22}$cm$^{-2}$)      &   0.53$\pm$0.06            &    0.47$\pm$0.04           &   0.48$\pm$0.02            &  0.49$\pm$0.03          &  0.47$\pm$0.03              &  0.52$_{-0.07}^{+0.04}$                                             \\
\\                                                                                                                                                                                                                                                                \\
{\sc bbody}    &$kT_{\rm BB}$ (keV)                   &   1.85$\pm$0.11            &    2.12$\pm$0.05           &   2.05$\pm$0.02            &  2.04$\pm$0.03          &  2.05$\pm$0.01              &  2.39$\pm$0.08                                                      \\
               &Norm       (10$^{-3}$)                &   1.0$\pm$0.1              &    7.5$_{-0.1}^{+0.4}$     &   12.7$\pm$0.2             &  11.2$\pm$0.3           &  13.0$\pm$0.2               &  3.0$\pm$0.2                                                \\
               &Flux (10$^{-10}$ c.g.s)               &   0.89$_{-0.12}^{+0.16 }$  & 6.35$_{- 0.12}^{+ 0.57 }$  &  10.70$_{-0.10}^{+0.17}$   &  9.49$_{-0.07}^{+0.32}$ &  11.07$\pm$0.08             &  2.53$\pm$0.21                                                        \\
\\                                                                                                                                                                                                                                                       \\
{\sc diskbb}   &$kT_{\rm disk}$ (keV)                 &   0.20$\pm$0.02            &    0.91$\pm$0.03           &   0.99$\pm$0.02            &  0.96$\pm$0.02          &   0.95$\pm$0.01             &   0.24$\pm$0.03                                              \\
               &Norm                                  &   14809$_{-7796}^{+15858}$ &    54$_{-11}^{+17}$        &   66$_{-7}^{+2}$            &  60$\pm$8              &   84$\pm$5                  &   7096$_{-4498}^{+5687}$                                                   \\
               &Flux (10$^{-10}$ c.g.s)               &    8.04$_{-1.12}^{+2.19 }$ &   8.59$_{- 0.67}^{+ 1.05 }$& 13.71$_{- 0.30 }^{+  0.52}$ & 11.15$_{-0.27}^{+0.77}$& 14.57$_{-0.27}^{+0.48}$     &  2.75$\pm$1.79                                             \\
\\                                                                                                                                                                                                                                                          \\
{\sc relxillCp}&$\beta$                               &   1.8$\pm$0.8              &    2.0$\pm$0.2             &   2.7$\pm$0.5              &  2.2$\pm$0.1            &   2.1$\pm$0.1               &   1.7$\pm$0.4                                                                  \\
               &$R_{\rm in}$ ($R_{\rm g}$)            &   8.37$_{-3.25*}^{+40.31}$ &    5.12$_{-0*}^{+3.28}$    &   15.02$_{-5.01}^{+5.35}$  &  5.12$_{-0*}^{+1.04}$   &   5.31$_{-0.19*}^{+1.44}$   &   5.60$_{-0.48*}^{+5.93}$                                     \\
               &$\Gamma$                              &   1.83$\pm$0.01            &    2.16$\pm$0.05           &   2.24$\pm$0.04            &  2.30$\pm$0.07          &   2.26$\pm$0.04             &   1.89$\pm$0.02                                                                   \\
               &log($\xi$)                            &   3.62$_{-0.13}^{+0.08}$   &    3.81$_{-0.10}^{+0.06}$  &   3.70$_{-0.06}^{+0.16}$   &  3.55$_{-0.21}^{+0.19}$ &   3.33$_{-0.06}^{+0.16}$    &   3.64$_{-0.04}^{+0.03}$                                \\
               &$kT_{e}$ (keV)                       &   18$\pm$2                 &    10$_{-1}^{+3}$          &   10.0$_{-0.4}^{+1.2}$      &  10$_{-1}^{+7}$        &   10$\pm$1                  &   42$_{-14}^{+29}$                                                              \\
               &R$_{refl}$                            &   1.8$_{-0.8}^{+1.3}$      &    10$_{-3}^{+0*}$         &   10$_{-1}^{+0*}$           &  10$_{-3}^{+0*}$       &   10$_{-1}^{+0*}$           &   10$_{-5}^{+0*}$                                                         \\
               &Norm (10$^{-4}$)                      &   16.94$_{-5.46}^{+4.30}$  &    5.45$_{-0.94}^{+ 1.11}$ &   5.63$_{-0.22}^{+1.20}$   &  4.59$_{-0.58}^{+1.38}$ &   4.64$_{-0.56}^{+0.77}$    &   7.06$_{-0.24}^{+2.18}$                                  \\
               &Flux (10$^{-10}$ c.g.s)               &  26.21$_{-0.76}^{+0.90 }$  &   28.13$_{-6.21}^{+ 1.88}$ & 29.80$_{- 2.40 }^{+ 6.01}$ & 26.15$_{-6.86}^{+1.35}$ &  21.22$_{-1.75}^{+2.19}$    &   35.15$_{-0.84}^{+0.58}$                           \\

\hline
               &$\chi^2_\nu$($\chi^2/dof)$   &   0.97 (171/177)  &   0.99 (156/157) &  1.24 (193/156)  & 0.85 (134/157)  &  1.25 (193/155)  & 1.20 (215/179)   \\
               &Total flux (10$^{-10}$ c.g.s)  & 30.81$_{-2.95}^{+3.98 }$  &  42.22$\pm$4.03  & 52.98$_{- 1.41 }^{+  1.94  }$ & 42.88$_{-4.01}^{+3.25}$ & 47.54$_{-1.08}^{+1.84}$ & 40.92$_{-3.04}^{+2.28}$ \\
\hline                                                                                                                
\hline
\end{tabular} }
\medskip  
\small
Note: We give the unabsorbed flux (erg cm$^{-2}$ s$^{-1}$) in the energy range 0.1-100 keV. All errors in the Tables are at the 90\%\ confidence level unless otherwise indicated. An asterisk symbol means that the error is pegged at the hard limit of the parameter range.
\label{relxill}
\end{table*}

\begin{table*}
\centering
\caption{Best-fitting results for the fit to the X-ray spectra of 4U 1636--53 with the reflection model {\sc bbody+diskbb+relxilllp}.}

\resizebox{\textwidth}{42mm}{

\begin{tabular}{cccccccc}
\hline
\hline
Model Comp   &   Parameter     & X1       & X2       & X3       & X4       & X5        & X6    \\
\hline

{\sc Tbabs}    &$N_{\rm H}$ (10$^{22}$cm$^{-2}$)      &      0.50$\pm$0.06             &   0.36$\pm$0.04           &  0.45$\pm$0.03          &  0.42$\pm$0.06             &     0.42$\pm$0.05            &   0.48$\pm$0.07                                                      \\
\\                                                                                                                                                                                    
{\sc bbody}    &$kT_{\rm BB}$ (keV)                   &      1.86$\pm$0.19             &   2.08$\pm$0.05           &  2.06$\pm$0.03          &  2.05$\pm$0.05             &     2.06$\pm$0.02            &   2.48$\pm$0.11                                                       \\
               &Norm       (10$^{-3}$)                &      0.6$\pm$0.2               &   7.0$\pm$0.4             &  12.0$\pm$0.3           &  11.1$\pm$0.6              &     12.6$_{-1.1}^{+0.2}$     &   2.4$\pm$0.3                                                         \\
               &Flux (10$^{-10}$ c.g.s)               &   0.55$_{-0.17}^{+0.20 }$      & 5.88$_{- 0.22}^{+0.35}$   & 10.23$_{-0.17}^{+0.25}$ &  9.54$\pm$0.42              &   10.23$_{-0.50}^{+0.34}$   &   2.01$_{-  0.24  }^{+  0.12  }$                                        \\                                              
\\                                                                                                                                                                               
{\sc diskbb}   &$kT_{\rm disk}$ (keV)                 &      0.22$\pm$0.02             &   0.85$\pm$0.03           &  0.97$\pm$0.02          &  0.94$\pm$0.04             &     0.94$\pm$0.01            &   0.26$\pm$0.03                                                       \\
               &Norm                                  &      8810$_{-3774}^{+9014}$    &   77$\pm$13               &  68$\pm$8               &  65$\pm$10                 &     85$\pm$5                 &   4448$_{-2872}^{+4025}$                                                \\
               &Flux (10$^{-10}$ c.g.s)               &     6.07$_{-0.92}^{+2.25 }$    &  8.57$_{- 0.62}^{+1.05}$  &  13.39$\pm$0.48         &  10.97$_{-0.49}^{+1.12}$   &   14.71$\pm$0.53             &   2.73$\pm$1.37                                                       \\
\\                                                                                                                                                                                 
{\sc relxilllp}&$h$ ($R_{\rm g}$)                     &      40$_{-15}^{+55}$          &   23$_{-7}^{+14}$         &  24$_{-9}^{+7}$         &  21$_{-7}^{+20}$           &     18$\pm$4                 &   33$_{-11}^{+18}$                                                  \\
               &$R_{\rm in}$ ($R_{\rm g}$)            &      5.12$_{-0*}^{+26.59}$     &   5.12$_{-0*}^{+4.85}$    &  5.12$_{-0*}^{+9.66}$   &  5.12$_{-0*}^{+2.23}$      &     5.12$_{-0*}^{+0.80}$     &   5.12$_{-0*}^{+6.13}$                                                       \\
               &$\Gamma$                              &      1.70$\pm$0.05             &   1.80$\pm$0.11           &  1.99$\pm$0.06          &  2.07$_{-0.17}^{+0.28}$    &     2.00$_{-0.50}^{+0.04}$   &   1.78$\pm$0.03                                                    \\
               &log($\xi$)                            &      3.56$_{-0.12}^{+0.04}$    &   3.70$_{-0.09}^{+0.06}$  &  3.67$_{-0.11}^{+0.07}$ &  3.66$_{-0.16}^{+0.10}$    &     3.44$\pm$0.11            &   3.57$_{-0.06}^{+0.03}$                                              \\
               &E$_{cut}$ (keV)                       &      60$\pm$14                 &   19$_{-2}^{+5}$          &  19$\pm$2               &  23$_{-4}^{+56}$           &     19$_{-7}^{+2}$           &   77$\pm$20                                                        \\
               &R$_{refl}$                            &      1.8$_{-0.7}^{+2.3}$       &   5.0$_{-2.5}^{+5*}$       &  10.0$_{-2.4}^{+0*}$    &  10.0$_{-6.2}^{+0*}$       &     10.0$_{-3.0}^{+0*}$      &   6.5$_{-4.0}^{+3.5*}$                                               \\
               &Norm (10$^{-4}$)                      &      17.49$_{-6.97}^{+3.35}$   &   7.18$_{-3.55}^{+5.54}$  &  5.05$\pm$0.54          &  3.64$_{-0.82}^{+3.60}$    &     3.92$_{-0.96}^{+1.48}$   &   9.60$_{-2.88}^{+8.70}$                                              \\
               &Flux (10$^{-10}$ c.g.s)               &      26.09$\pm$0.85            &  21.45$_{-4.79}^{+2.53}$  & 24.87$_{-5.55}^{+2.02}$ &  17.16$_{-7.98}^{+2.37}$   &  15.54$_{-3.10}^{+3.36}$     &   32.99$_{-  1.19  }^{+  2.28  }$     \\

\hline
               &$\chi^2_\nu$($\chi^2/dof$)     & 0.93 (165/177)& 0.98 (154/157) & 1.19 (186/156) & 0.93 (146/157) &  1.28 (198/155) &  1.20 (214/179)    \\
               &Total flux (10$^{-10}$ c.g.s)  & 32.47$_{-1.95}^{+2.78 }$ & 36.14$_{- 2.74}^{+ 1.81 }$ & 48.82$_{- 3.06}^{+1.44}$  & 37.15$_{-6.10}^{+3.53}$ & 40.08$_{-1.51}^{+3.54}$ & 37.99$_{-  2.24  }^{+  1.47  }$    \\
\hline                                                                                                                
\hline
\end{tabular}

}

\medskip  
\label{relxilllp}
\end{table*}

\begin{table*}
\centering
\caption{Best-fitting results for the fit to the X-ray spectra of 4U 1636--53 with the reflection model {\sc bbody+diskbb+relxilllpCp}.}
\resizebox{\textwidth}{42mm}{

\begin{tabular}{cccccccc}
\hline
\hline
Model Comp   &   Parameter     & X1       & X2       & X3       & X4       & X5        & X6    \\
\hline

{\sc Tbabs}    &$N_{\rm H}$ (10$^{22}$cm$^{-2}$)      &      0.56$\pm$0.05              &    0.44$\pm$0.04             &    0.49$\pm$0.02            &   0.47$\pm$0.05            &  0.46$\pm$0.03             &   0.51$_{-0.08}^{+0.04}$                                                 \\
\\                                                                                                                                                                                                                                                                                   
{\sc bbody}    &$kT_{\rm BB}$ (keV)                   &      2.00$\pm$0.10              &    2.11$\pm$0.05             &    2.06$\pm$0.03            &   2.05$\pm$0.03            &  2.05$\pm$0.02             &   2.40$\pm$0.08                                                           \\
               &Norm       (10$^{-3}$)                &      1.0$_{-0.1}^{+0.2}$        &    7.8$\pm$0.3               &    12.6$\pm$0.2             &   11.2$\pm$0.3             &  13.0$\pm$0.2              &   3.0$\pm$0.3                                                             \\
               &Flux (10$^{-10}$ c.g.s)               &     0.88$_{-0.11}^{+0.14 }$    &   6.57$_{- 0.14}^{+ 0.32 }$   &  10.76$_{-0.18}^{+0.08}$    &   9.58$\pm$0.24            &  11.02$_{-0.17}^{+0.25  }$ &   2.54$_{-  0.21  }^{+  0.30  }$                                             \\                                              
\\                                                                                                                                                                                                                                                                                   
{\sc diskbb}   &$kT_{\rm disk}$ (keV)                 &      0.19$\pm$0.01              &    0.90$\pm$0.03             &    1.00$\pm$0.02            &   0.96$\pm$0.02            &  0.95$\pm$0.01             &   0.23$\pm$0.02                                                           \\
               &Norm                                  &      21452$_{-7641}^{+11648}$   &    62$\pm$13                 &    62$\pm$6                 &   60$\pm$7                 &  84$\pm$6                  &   7146$_{-4635}^{+6578}$                                                    \\
               &Flux (10$^{-10}$ c.g.s)               &     8.44$_{-1.68}^{+1.83 }$   &    8.72$_{- 0.50}^{+ 1.16 }$   &  13.78$_{- 0.32}^{+0.60}$   &  11.28$_{-  0.56}^{+0.70}$ & 15.04$_{-0.71}^{+0.32  }$  &   2.90$_{-  1.60  }^{+  2.51  }$                                            \\
\\                                                                                                                                                                                                                                                                                   
{\sc relxilllpCp} &$h$ ($R_{\rm g}$)                   &      100$_{-92}^{+0*}$          &    27$_{-10}^{+16}$          &    27$_{-15}^{+6}$          &   21$_{-5}^{+12}$          &  22$_{-4}^{+6}$            &   38$_{-13}^{+30}$                                                      \\
               &$R_{\rm in}$ ($R_{\rm g}$)            &      50.00$_{-44.88*}^{+0*}$    &    5.12$_{-0*}^{+5.99}$      &    5.66$_{-0.54}^{+13.41}$  &   5.12$_{-0*}^{+1.64}$     &  5.12$_{-0*}^{+1.25}$      &   5.12$_{-0*}^{+22.16}$                                                          \\
               &$\Gamma$                              &      1.83$\pm$0.01              &    2.13$\pm$0.06             &    2.26$\pm$0.04            &   2.29$\pm$0.10            &  2.26$\pm$0.05             &   1.89$\pm$0.02                                                        \\
               &log($\xi$)                            &      3.72$_{-0.20}^{+0.07}$     &    3.81$_{-0.10}^{+0.08}$    &    3.70$_{-0.06}^{+0.05}$   &   3.68$_{-0.15}^{+0.13}$   &  3.47$_{-0.14}^{+0.16}$    &   3.64$_{-0.05}^{+0.03}$                                                  \\
               &$kT_{e}$ (keV)                       &      24$_{-5}^{+10}$            &    10$_{-2}^{+3}$            &    10$\pm$1                 &   10$_{-2}^{+51}$          &  10$\pm$1                  &   41$_{-14}^{+46}$                                                     \\
               &R$_{refl}$                            &      2.9$\pm$1.9                &    10.0$_{-6.0}^{+0*}$       &    10.0$_{-1.7}^{+0*}$      &   10.0$_{-5.3}^{+0*}$      &  10.0$_{-1.7}^{+0*}$       &   8.0$_{-4.3}^{+2.0*}$                                                   \\
               &Norm (10$^{-4}$)                      &      13.36$_{-4.61}^{+10.99}$   &    4.94$_{-0.88}^{+1.57}$    &    5.92$_{-0.63}^{+1.01}$   &   4.40$_{-0.80}^{+1.12}$   &  4.48$_{-0.34}^{+0.83}$    &   8.31$\pm$1.02                                                           \\
               &Flux (10$^{-10}$ c.g.s)               &     27.11$_{-1.71}^{+0.57 }$    &  27.55$_{- 6.37}^{+ 2.92 }$  &  28.28$_{- 1.76 }^{+2.60}$  &  20.85$_{-  5.50}^{+4.18}$ & 21.16$_{-3.11 }^{+4.39  }$ &   35.15$_{-  0.87  }^{+  0.93  }$                                   \\

\hline
               &$\chi^2_\nu$($\chi^2/dof)$     &  0.98 (173/177) & 1.01 (158/157) & 1.23 (192/156) &  0.92 (144/157) & 1.33 (206/155)  &  1.22 (219/179)   \\
               &Total flux (10$^{-10}$ c.g.s)  & 33.23$_{-2.75}^{+4.03 }$&  43.22$_{- 5.91}^{+ 2.09 }$  &  53.73$_{- 1.96 }^{+  1.28  }$  & 36.82$_{-  3.26  }^{+  4.73  }$  & 49.35$_{-   3.39  }^{+  0.90  }$ & 40.50$_{-  4.05  }^{+  2.71  }$    \\
\hline                                                                                                                
\hline
\end{tabular}  }

\medskip  
\label{relxilllpcp}
\end{table*}

\begin{table*}

\small

\centering
\caption{Best-fitting results for the fit to the X-ray spectra of 4U 1636--53 with the reflection model {\sc bbody+diskbb+cutoffpl+relconv$\times$bbrefl}.}

\begin{tabular}{cccccc}
\hline
\hline
Model Comp   &   Parameter       & X2       & X3       & X4       & X5       \\
\hline

{\sc Tbabs}    &$N_{\rm H}$ (10$^{22}$cm$^{-2}$)      &     0.27$_{-0.05}^{+0.15}$   &      0.40$_{-0.14}^{+0.05}$   &    0.51$_{-0.07}^{+0.03}$   &     0.47$_{-0.09}^{+0.04}$                             \\
\\                                                                                                                                  
{\sc bbody}    &$kT_{\rm BB}$ (keV)                   &     1.91$_{-0.13}^{+0.23}$   &      2.09$_{-0.13}^{+0.09}$   &    2.09$\pm$0.05            &     2.12$\pm$0.04                                      \\
               &Norm       (10$^{-3}$)                &     3.1$_{-1.5}^{+1.2}$      &      4.3$_{-2.6}^{+2.0}$      &    8.5$_{-1.5}^{+1.0}$      &       8.8$\pm$1.2                                      \\
               &Flux (10$^{-10}$ c.g.s)               &   2.64$_{-1.32}^{+1.04}$     &       3.37$_{-1.32}^{+1.47}$  &    6.99$_{-1.38}^{+0.93}$   &            7.17$_{-1.30}^{+0.97}$                                              \\
\\                                                                                                                                   
{\sc diskbb}   &$kT_{\rm disk}$ (keV)                 &     0.78$\pm$0.05            &      0.91$\pm$0.12            &    0.98$_{-0.06}^{+0.03}$   &     0.97$_{-0.05}^{+0.01}$                             \\
               &Norm                                  &     133$_{-59}^{+21}$        &      77$_{-18}^{+39}$         &    55$\pm$8                 &     76$\pm$8                                         \\
               &Flux (10$^{-10}$ c.g.s)               &     10.70$_{-2.10}^{+1.65}$  &   13.94$_{-1.40}^{+3.78}$     &   11.26$_{-2.58}^{+0.80}$   &           15.05$_{-2.18}^{+1.34}$                                              \\
\\                                                                                                                                   
{\sc cutoffpl} &$\Gamma$                              &     1.20$_{-0*}^{+0.76}$      &      1.65$_{-0.45*}^{+0.78}$   &    2.5$_{-0.5}^{+0.1}$      &     2.50$_{-0.60}^{+0.01}$                             \\
               &E$_{cut}$ (keV)                       &     8.9$_{-0.8}^{+8.8}$      &      9$_{-3}^{+91*}$           &    100$_{-87}^{+0*}$        &     100$_{-88}^{+0*}$                                  \\
               &Norm                                  &     0.13$_{-0.05}^{+0.16}$   &      0.34$_{-0.22}^{+0.07}$   &    0.31$\pm$0.04            &     0.28$\pm$0.06                                      \\
               &Flux (10$^{-10}$ c.g.s)               &  13.74$_{-4.75}^{+6.79}$     &    14.65$_{-9.94}^{+5.12}$    &   24.17$_{-12.37}^{+2.94}$  &           20.76$_{-14.05}^{+3.80}$                                                     \\
\\                                                                                                                                  
{\sc relconv}  &$\beta$                               &     2.2$\pm$0.3              &      2.3$\pm$0.2              &    2.3$\pm$0.2              &     2.4$\pm$0.1                                      \\
{\sc bbrefl}   &log($\xi$)                            &     3.18$_{-0.81}^{+0.34}$   &      3.25$_{-1.00}^{+0.27}$   &    3.11$_{-0.26}^{+0.18}$   &     3.17$_{-0.11}^{+0.13}$                             \\
               &Norm                                  &     2.8$_{-0.7}^{+13.6}$     &      5.0$_{-0.8}^{+6.0}$      &    3.9$_{-0.8}^{+2.5}$      &     4.9$_{-0.6}^{+1.1}$                             \\
               &Flux (10$^{-10}$ c.g.s)               &   2.85$_{-1.02}^{+5.15}$     &     6.90$_{-3.78}^{+6.44}$    &    3.73$_{-1.57}^{+1.95}$   &     5.34$_{-2.07}^{+2.60}$                                                   \\
        
\hline
               &$\chi^2_\nu$($\chi^2/dof)$ &  0.94 (149/158)   &  1.01 (158/157)   &  0.78 (124/158)   &  1.06 (166/156)          \\
               &Total flux (10$^{-10}$ c.g.s)  &  29.89$_{-0.45}^{+4.02}$     &   38.09$_{-0.52}^{+4.40}$   &    42.46$_{-10.68}^{+8.30}$   &     40.62$_{-5.56}^{+11.40}$  \\
\hline                                                                                                                
\hline
\end{tabular} 
\medskip  
\\
\label{bbrefl}
\end{table*}

\begin{table*}

\small
\centering
\caption{Best-fitting results for the fit to the X-ray spectra of 4U 1636--53 with the reflection model {\sc diskbb+cutoffpl+relxillNS}.}

\begin{tabular}{cccccc}
\hline
\hline
Model Comp   &   Parameter       & X2       & X3       & X4       & X5       \\
\hline

{\sc Tbabs}    &$N_{\rm H}$ (10$^{22}$cm$^{-2}$)      &     0.27$\pm$0.08          &  0.49$_{-0.11}^{+0.07}$      &     0.52$_{-0.06}^{+0.03}$     &     0.53$_{-0.08}^{+0.03}$                        \\
\\                                                                                                                                                                                             
{\sc diskbb}   &$kT_{\rm disk}$ (keV)                 &     0.76$\pm$0.03          &  0.96$_{-0.14}^{+0.04}$      &     0.96$\pm$0.05              &     0.97$_{-0.05}^{+0.02}$                        \\
               &Norm                                  &     146$_{-41}^{+19}$      &  64$_{-5}^{+25}$             &     53$_{-8}^{+13}$            &     71$\pm$8                                    \\
               &Flux (10$^{-10}$ c.g.s)               & 10.47$_{- 2.51}^{+ 1.11 }$ & 11.58$_{- 2.77}^{+4.67}$     &  9.70$_{-  1.37  }^{+1.17}$    &   13.35$_{-   1.46  }^{+  1.69  }$                                              \\
\\                                                                                                                             
{\sc cutoffpl} &$\Gamma$                              &     1.20$_{-0*}^{+0.71}$   &  2.12$_{-0.62}^{+0.54}$      &     2.39$_{-0.41}^{+0.23}$     &     2.56$_{-0.54}^{+0.07}$                          \\
               &E$_{cut}$ (keV)                       &     8.7$_{-0.6}^{+16.4}$   &  15$_{-8}^{+85*}$            &     36$_{-23}^{+64*}$          &     98$_{-85}^{+2*}$                               \\
               &Norm                                  &     0.14$_{-0.02}^{+0.06}$ &  0.42$\pm$0.06               &     0.34$\pm$0.04              &     0.35$\pm$0.05                                 \\
               &Flux (10$^{-10}$ c.g.s)               & 14.08$_{- 3.08}^{+ 4.60 }$ &    22.73$\pm$3.48            &  20.50$_{-  5.56}^{+  7.56  }$ &   18.97$_{-   4.41  }^{+ 11.46  }$                                                       \\
\\                                                                                                                         
{\sc relxillNS}  &$\beta$                             &     2.2$\pm$0.3            &  2.3$\pm$0.2                 &     2.3$\pm$0.2                &     2.4$\pm$0.1                                                            \\
                 &$R_{\rm in}$ ($R_{\rm g}$)          &     5.12$_{-0*}^{+3.38}$   &  5.14$_{-0.02*}^{+1.78}$     &     5.12$_{-0*}^{+0.70}$       &     5.12$_{-0*}^{+0.49}$                                  \\
                 &$kT_{\rm BB}$ (keV)                 &     1.85$_{-0.37}^{+0.14}$ &  2.14$_{-0.22}^{+0.07}$      &     2.04$_{-0.22}^{+0.08}$     &     2.10$\pm$0.04            \\
                 &log($\xi$)                          &     2.93$_{-1.56}^{+0.52}$ &  3.35$_{-0.63}^{+0.12}$      &     2.81$_{-1.22}^{+0.43}$     &     3.08$_{-0.23}^{+0.18}$                          \\
                 &R$_{refl}$                          &     2.6$_{-1.0}^{+58.6}$   &  8.6$_{-5.7}^{+3.2}$         &     1.2$\pm$0.3                &     1.6$\pm$0.5                                                       \\
                 &Norm (10$^{-4}$)                    &     3.1$_{-2.3}^{+1.4}$    &  2.6$_{-0.6}^{+2.3}$         &     8.5$_{-1.6}^{+1.2}$        &     8.4$_{-1.7}^{+1.1}$                               \\
                 &Flux (10$^{-10}$ c.g.s)             & 5.32$_{- 0.45}^{+ 1.01 }$  &  9.65$_{-1.83}^{+3.93  }$    &    9.76$_{-1.63}^{+ 1.01  }$   &     11.72$_{-   1.11  }^{+  0.88  }$                     \\
        
\hline
               &$\chi^2_\nu$($\chi^2/dof)$     &  0.97 (152/157) & 0.99 (155/156) & 0.80 (126/157) & 1.05 (162/155)                                                      \\
               &Total flux (10$^{-10}$ c.g.s)  &   29.89$_{- 0.66}^{+ 3.26 }$ & 43.61$_{- 2.37 }^{+ 11.25  }$ &  36.69$_{-  6.91  }^{+  8.45  }$   &   44.01$_{-   7.78  }^{+ 10.32  }$                                        \\
\hline                                                                                                                
\hline
\end{tabular} 
\medskip  
\label{relxillns}
\end{table*}

\begin{figure*}
\center
\includegraphics[width=0.50\textwidth]{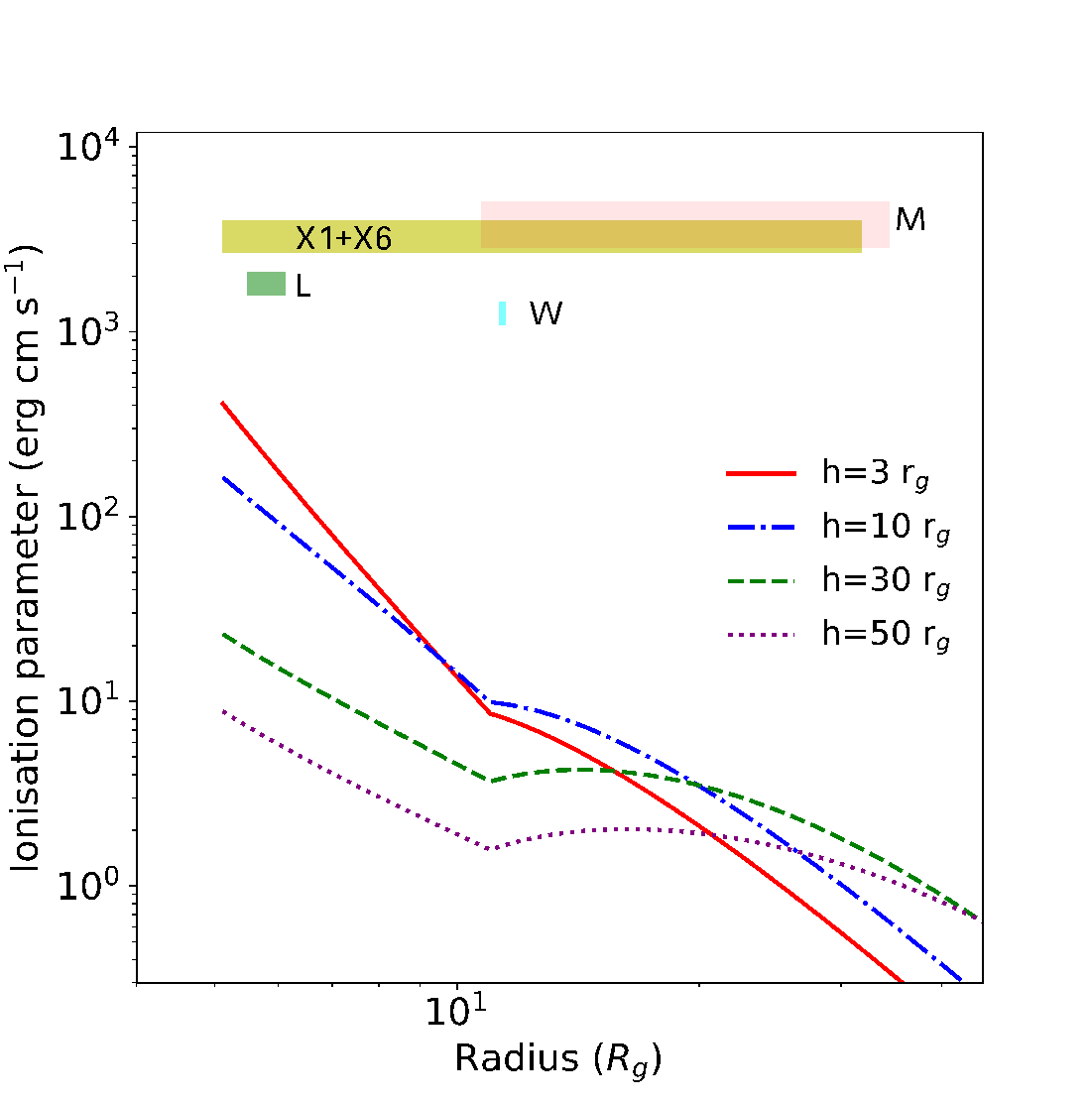}
\caption{Predicted ionization profiles of an accretion disk illuminated by a lamppost X-ray source located at different heights above the central source. We applied Eq.  (1) from   \citet{ballantyne17} for the curve generation. We assumed $\eta$=0.2, $\alpha$=0.3, $\lambda$=0.2, and $f$=0.45, and set the $(R_R, R_z, R_T)$ factors at $r< 11$ $R_{g}$ to their values at $r= 11$ $R_{g}$ to avoid an unphysical break in $\xi(r,h)$. In the plot,  the predicted ionization curves are compared with the ionization parameter measured from the observations in the hard spectral state in \citet{ludlam17} (L), \citet{wyn17} (W), \citet{mondal21} (M), and this work (X1+X6), respectively.}
\label{compa}
\end{figure*}

The parameter reflection fraction, R$_{refl}$, is very high in most of the fits, which is likely due to the complicated reflection model that we used. To eliminate the possible influence of this high reflection fraction, we further fit the spectra with the  {\sc relxilllp} model, but with the reflection fraction parameter fixed at certain values. According to the document of the {\sc relxill} model,\footnote{http://www.sternwarte.uni-erlangen.de/~dauser/research/relxill/} the reflection fraction is expected to be larger than 1 since photons from the corona are preferentially bent  toward the disk by the strong gravity. Moreover, simulation results indicate that the reflection fraction shows an increasing trend as the inner radius of the disk decreases \citep{dauser14}. Therefore, for the observations in the hard state, we fixed the reflection fraction to  unity since the inner radius is large, while we set it to be 2 for those soft observations as the disk generally moves to a position closer to the central neutron star in the soft state. The corresponding fitting result (Table \ref{fix}) shows that the variations in the parameters are not significant compared with those in Table \ref{relxilllp}, suggesting that the high reflection fraction in the fits only has very limited influence on the conclusion in this work.

\section{Discussion} 
In this work we investigated the reflection off the disk in the neutron star low-mass X-ray binary 4U 1636--53 with the XMM-{\it Newton} and the simultaneous {\it RXTE} observations. We found that the spectra in the soft state can be well described if the disk is illuminated by either a nonthermal component from the corona or a thermal component from the neutron star surface. Additionally, the emissivity index derived in both the neutron star illumination models and the corona illumination models are relatively small, $\sim$ 2.2, lower than the value of 3 predicted from theory. We also found that the accretion disk becomes less ionized if the illumination comes from the neutron star surface. Finally, there is likely a relationship  between the height of the corona and the normalization of the disk radiation, which is connected to the inner radius of the disk. 

The reflection spectra of 4U 1636--53 have already been studied by previous authors. \citet{ludlam17} studied one NuSTAR observation in the hard state with the  {\sc relxill} model alone, and   found that the R$_{in}$ and log($\xi$) is around 1.03 $R_{ISCO}$ and 3.26, respectively. Later, \citet{wyn17} investigated its spectra properties in the soft, transitional, and hard states with three NuSTAR observations with a model consisting of a blackbody component and a {\sc relxilllp} component. The derived ionization parameter log($\xi$) decreases from the soft state to the hard state, with the corresponding value of 4.4$\pm$0.03, 3.4$\pm$0.03, and 3.1$\pm$0.06. Additionally, the derived height of the corona in these observations is very small, $\sim$ 2-3 $R_{g}$. Recently, \citet{mondal21} analyzed one NuSTAR observation in the hard state with the  {\sc nthcomp+relxillcp} model. They found that the R$_{in}$ and the log($\xi$) of the accretion disk is 4.8$_{-2.7}^{+1.9}$ $R_{ISCO}$ and 3.58$\pm$0.12 $R_{ISCO}$, respectively. 

\subsection{Illumination from the neutron star}
In a neutron star system, the reflection scenario is more complicated than in a black hole system since the neutron star could also illuminate and ionize the disk. The neutron star illumination becomes significant when the disk comes close to the neutron star in the soft state. Theoretically, \citet{ballan04} developed a model of X-ray reflection from a constant-density slab illuminated by a blackbody component from a neutron star X-ray burst. Their calculations predicted a spectrum composed of a prominent iron line, rich soft X-ray lines, and a strong free–free continuum, consistent with the spectral features in observations. Later, \citet{wilkin18} explored the illumination of the accretion disk by radiation emitted from the neutron star surface through a general-relativistic ray-tracing method. Theoretical emissivity profiles were calculated assuming that the disk is illuminated by hotspots, bands of emission, and the entirety of the spherical star surface, respectively. Interestingly, the emissivity profile in all three cases could be well represented by a single power law slightly steeper than r$^{-3}$. Recently, \citet{garcia22} presented a new reflection model to describe the X-ray radiation reprocessed in the accretion disk around the neutron star, with a single-temperature blackbody illumination coming from the surface of the star or its boundary layer. In general, their model shows good consistency with the earlier model {\sc bbrefl};   the difference is mainly in the shape of the iron profile and the continuum at softer energies. A narrower Fe K emission line is obtained in their calculation. 

Observationally, illumination from the neutron star and its boundary layer has been explored in some previous works \citep[e.g.,][]{daa10,cackett10,sanna13,chiang16,malu20,ludlam17,ludlam18,ludlam20,ludlam22}. \citet{daa10} analyzed two XMM-{\it Newton} observations of the bright atoll source 4U 1705--44, and found that it is reasonable that the reflection component in the soft state comes from hard X-ray thermal irradiation from the neutron star boundary layer. \citet{cackett10} systematically studied ten neutron star LMXBs, and found that in most cases their spectra could be well fitted if the disk is illuminated by a blackbody component from the neutron star. \citet{ludlam20} fitted one simultaneous NICER and NuSTAR spectrum of the neutron star binary 4U 1735--44 well with the neutron star illumination model {\sc relxillNS}, with the inner radius of the disk and the maximum extent of the boundary layer being accurately constrained. The fitting results in this work indicate that the soft observations were well fitted with the neutron star illumination models, suggesting that the accretion disk could be illuminated by the neutron star surface in 4U 1636--53. 

Interestingly, for these soft observations the emissivity index in both the neutron star illumination model {\sc bbrefl} and the corona illumination model {\sc relxillCp} are relatively low, $\sim$ 2.3 and $\sim$ 2.2, significantly lower than the typical value of 3 from theoretical simulations. In the case of neutron star illumination, \citet{wilkin18} shows that the emissivity index is always $\sim$ 3 regardless of whether the illuminating blackbody component comes from a hotspot region, a belt region, or the surface of the neutron star. As a comparison, for the corona illumination in active galactic nuclei, \citet{wilkins11} found that in a form of a single power law, the derived emissivity index is also similar: $\sim$ 3.3.

Recently, \citet{ludlam22} analyzed the accreting neutron star low-mass X-ray binary Cygnus X--2 using the neutron star illuminating model {\sc relxillNS} and the corona illuminating model {\sc rfxconv}, with the derived emissivity index being $\sim$ 2 and $\sim$ 1.5, respectively. Moreover, similar emissivity indices were found in the reflection spectra of the neutron-star LMXB GX 13+1 using {\em NuSTAR} data by \citet{saavedra2023}. Their derived emissivities are relatively small, with the values close to the predicted shallower illumination profile from an extended corona around a slowly spinning compact object \citep{kinch16,kinch19}. \citet{kinch16} show that the simulated Fe K$\alpha$ emissivity is $\varpropto$ r$^{-2}$, which is proposed to be a direct consequence of the extended coronal emission as the emissivity index becomes 3 if the corona is treated as a point source at height z$\ll$r. For the same reason, one thing possibily responsible for the low emissivity index in 4U 1636--53 could be that the accretion disk is illuminated by the neutron star and the corona simultaneously. As a consequence, the neutron star and the corona actually constitute an extended illuminator, with the degree of the extension connected to the distance between them. So in this case they could illuminate the accretion disk from different angles, which would lead to an  emissivity index of around 2, due to the shallower illumination profile in an extended coronal geometry.

Moreover, we note that the emissivity index tends to be lower (Table \ref{relxill}) in the hard-state observations when the height of the corona is greater, consistent with the simulation result. \citet{dauser13} modeled the illumination from the primary X-ray source for different geometries of the primary source. They predicted the emissivity profile for a moving primary source at different heights, with the speed at 0, 0.5c, and 0.9c. For a given height, the emissivity index becomes lower as the moving velocity increases, due to the Doppler boosting effect. Meanwhile, the emissivity index shows a clearly decreasing trend as the height of the primary source increases, in agreement with the result here that the emissivity index tends to be smaller for a greater corona height.

The neutron star illumination also has an important influence on the ionization state of the accretion disk. A significant difference between the ionization parameter in the neutron star illuminating case and the corona illuminating case is present in all four soft observations. One possible reason could be that the blackbody illuminating component has fewer high-energy photons compared with the power-law illuminating component, and thus the incident flux to the disk is lower, leading to a less ionized accretion disk in the neutron star illuminating case. 

Since the illuminating source has a significant influence on  the physical properties of the accretion disk, it is necessary to apply a more self-consistent physical model including illumination from both the corona and the neutron-star to determine the physics of the accretion disk in a more accurate way. In this case, the strength of the two types of illumination, together with their variations, could be a good indicator of how the geometry and the physics of the accretion system change as the source walks between different spectral states. However, the quality of the current data could hardly afford such a complicated reflection model; therefore, observations with long enough exposure times are needed in the future.

\subsection{ionization of the accretion disk} 

The ionization parameter derived in this work in the corona illuminating model indicates that the accretion disk was highly ionized in 4U 1636--53. \citet{ballantyne17} investigated the ionization parameter at different radii when the disk is illuminated by the corona above the black hole symmetric axis. They developed a formula to describe the relation between the height of the corona and the ionization of the disk:    

\begin{equation}
\begin{aligned}
\xi(r,h)= & (5.44\times 10^{10}) \left ({\eta \over 0.1} \right )^{-2} \left (
  {\alpha \over 0.1} \right) \lambda^3 \left ( {r \over r_g} \right )^{-3/2} R_z^{-2} R_T^{-1} \\
  & \times R_R^3 f(1-f)^3 F(r,h) g_{lp}^2 \mathcal{A}^{-1} \mathrm{erg\ cm\ s^{-1}}.
\label{eq:newxi}
\end{aligned}
\end{equation}
Here $\xi(r,h)$ is the ionization parameter of the disk at a radius of $r$; $h$ is the height of the corona in the lamppost geometry; $\eta$ is the radiative efficiency of the accretion; $\alpha$ is the viscosity parameter;  $\lambda=L_{bol}/L_{Edd}$ is the Eddington ratio; $f$ is the coronal dissipation fraction; and $R_R, R_z$, and $R_T$ are relativistic corrections to the Newtonian $\alpha$-disk equations \citep{krolik99}. The quantity $g_{lp}$=$\nu_{\mathrm{disk}}$/$\nu_{\mathrm{src}}$ is the measured frequency of a photon striking the disk divided by its frequency at the source \citep{dauser13}. The function $F(r,h)$ describes the shape of the irradiation pattern \citep{Fukumura07}, and $\mathcal{A}$ is the integral of $F(r,h)\times g_{\mathrm{lp}}^2$ over the entire disk, $\mathcal{A}=\int_{r_{\mathrm{in}}}^{r_{\mathrm{out}}} F(r,h) g_{\mathrm{lp}}^2 dS(r)$.

During the calculation, we set  $\eta$=0.2, $\alpha$=0.3 \citep{ballantyne17,penna13}, $\lambda$=0.2, and $f$=0.45 \citep{vasud07}. We found that the ionization curve breaks at around 11 $R_{g}$, which is a consequence of the divergence of $R_R$ and $R_T$ as the radius approaches the innermost stable circular orbit \citep{ballantyne17}. We then followed the same procedure as in \citet{ballantyne17} to fix it: the $(R_R, R_z, R_T)$ factors at $r< 11$ $R_{g}$ are fixed at their values at $r= 11$ $R_{g}$.

In Fig. \ref{compa} we show the ionization profile when the disk is illuminated by the corona at different heights. We compare these profiles with the results derived from observations in the hard spectral state:  X1 and X6 in this work; the NuSTAR observations in the work of \citet{ludlam17}, \citet{wyn17}, and \citet{mondal21}. We found that the ionization curves calculated from the formula do not reach the ranges derived from the  observations. All these predicted curves are located at   lower positions compared with the results from the observations. This is similar to the finding reported in previous works. \citet{lyu19} studied reflection properties in another neutron-star LMXB 4U 1728--34, and found that the ionization parameter from the fits is bigger than that predicted by the model. \citet{lyu19} proposed that this difference is likely due to the influence of the irradiation from the neutron star and its boundary layer, which could ionize the accretion disk, but is not included in the standard reflection model. Instead, here we only select the observations in the hard spectral state in Fig. \ref{compa}, so the illumination from the neutron star surface should be not very significant since the disk in the hard state is generally far from the central compact object. 

Moreover, it should be noted that the ionization difference between the observations and   theory could be even larger than that shown in the figure. On the one hand, the ionization parameter derived from the spectroscopy is actually the mean value of the whole disk, so the real ionization parameter in the innermost disk should be much larger than the averaged one shown in the plot, as the inner disk region receives more illumination than other parts. On the other hand, recent theoretical studies show that the corona likely moves outward at a relativistic speed \cite[e.g.,][]{beloborodov99}, and as a consequence the beaming effect likely reduces the illumination of the disk from the corona, leading to an even lower ionization profile.

The reason   for the ionization difference is still unknown at the current stage; however, we note that in the standard reflection model it is assumed that a lamppost X-ray source emits isotropically, which is likely not to be satisfied in a realistic context, due to the possible presence of anisotropic soft input photons \citep[e.g.,][]{Ghisell91,haardt91,haardt93,zhangwd19}. \citet{haardt93} investigated the angular distribution of the emitted power-law photons when the input thermal photons for the scattering are anisotropic. Their study showed that in the first scattering event there is a significant anisotropic distribution of hard radiation, with most of the power radiated back along the direction of the input seed photons. Later, \citet{zhangwd19} found that in the case of a corona above the disk, due to the anisotropic input of seed photons, the spectrum is more powerful in the direction of the seed photons. Consequently, the anisotropic corona could influence the reflection continuum and the Fe k$\alpha$ line flux. Specifically, this effect enables the accretion disk, working as a seed photon provider, to receive more illuminating luminosity from the corona, leading to a reflection fraction greater than unity. Therefore, for a less realistic isotropic corona in the standard reflection model, the accretion disk likely becomes less ionized if the illumination from the corona is underestimated.

\subsection{Height versus R$_{in}$} 
The distribution trend between the height of the corona and the normalization of the disk radiation suggests that, from the soft state to the hard state, the corona moves farther away from the neutron star as the accretion disk moves outward. This scenario is also supported  by the result reported in \citet{wyn17}. The height of the corona in their work is 2.3$\pm$0.2 R$_{g}$, 2.5$\pm$0.1 R$_{g}$, and 2.8$_{-0.3}^{+0.1}$ R$_{g}$ as the inner radius increases from $\sim$ 5.7 R$_{g}$ to 11.4 R$_{g}$ when the source evolves from the soft state to the hard state.

In the current stage the formation, structure, and dynamics of the corona are still poorly understood. Three types of  corona model have been proposed so far: (1) static corona with different geometries \citep[e.g., slab,][]{haardt94}; (2) base of a jet \citep{markoff01,markoff05}; (3) radiatively inefficient accretion flow (RIAF) \citep{ferreira06,narayan08}. In both the jet model and the RIAF model the corona is moving, consistent with the physical picture seen in relativistic magnetohydrodynamics (MHD) simulations \citep[e.g.,][]{beloborodov99}. 

The height of the corona is a key parameter for the ionization and the reflection off the disk. Theoretically, the corona is assumed to be located in a region of 3-100 $R_{g}$ \citep{ballantyne17} in a lamppost geometry. In the jet model it is assumed that the base of the jet is located within a few tens of gravitational radii \citep{markoff05}. These predicted ranges are consistent with the values obtained in this work. While the derived height in the work of \citet{wyn17} is relatively small, $\sim$ 2-3 $R_{g}$. \citet{wyn17} proposed that this small height may refer to the boundary layer between the disk and the neutron star surface as the illuminating hard X-rays in a neutron star system.

The finding that the corona is located far away from the central compact object in the hard state is consistent with the polarization results of Cygnus X-1 in \citet{chauvin18}. Their linear polarization measurements in the hard state reveal a low polarization fraction ($<$8.6\%) and the alignment of the polarization angle with the jet axis, which indicates that, for a lamppost geometry, the corona in the hard spectral state must be far away from the black hole so that the dominant emission is not significantly influenced by strong gravity.

The scenario that in the hard state the  corona is far from the compact object can be understood when considering the influence of the radiation pressure. In the model of \citet{beloborodov99}, the corona is the place where magnetic stress is transported and released in the form of bright flares. If the flaring plasma is composed of electron--positron pairs, then it will be accelerated and ejected away from the disk by the pressure of the reflected radiation. When the source is in the hard state, the reflection from the disk is stronger, thus pushing the plasma to a more distant location from the central neutron star. This scenario agrees well with the recent spectral analysis of the black hole X-ray binary MAXI J1820+070, which shows a decreasing trend (from $\sim$ 25 $R_{g}$ to $\sim$ 9 $R_{g}$) of the height of the corona in a hard-to-soft state transition process \citep{wang21}.

Alternatively, the finding can also be explained in the context of a jet base. As shown in \citet{ferreira06}, the canonical spectral states were determined by the variation in the transitional radius of the disk and the accretion rate. For the hard state, the transitional radius is big and the hard X-ray is expected to form at the jet base, which is located within a few tens of gravitational radii. While in the soft state, the transitional radius becomes smaller than the marginally stable orbit, hence the whole disk resembles a standard thin disk with no jet.  In this case, the weak hard X-rays likely originate from scattering by hybrid thermal--nonthermal electrons in active regions above the disk surface \citep{zdziarski04}. If these active regions are located close to the disk surface, then we   also see a decrease in the height of the corona as the source goes from the hard state to the soft state.

\section*{Acknowledgements}
 
We thank the anonymous referee for his/her careful reading of the manuscript and useful comments and suggestions. This research has made use of data obtained from the High Energy Astrophysics Science Archive Research Center (HEASARC), provided by NASA's Goddard Space Flight Center. This research made use of NASA's Astrophysics Data System. Lyu is supported by Hunan Education Department Foundation (grant No. 21A0096). HGW is supported by NSFC 12133004. FG is a CONICET researcher and acknowledges support from PIP 0113 and PIBAA 1275 (CONICET).

% bibliography data in biblio.bib
\bibliographystyle{aa}
\bibliography{biblio}

\begin{thebibliography}{77}
\expandafter\ifx\csname natexlab\endcsname\relax\def\natexlab#1{#1}\fi

\bibitem[{{Altamirano} {et~al.}(2008){Altamirano}, {van der Klis},
  {M{\'e}ndez}, {Jonker}, {Klein-Wolt}, \& {Lewin}}]{diego08}
{Altamirano}, D., {van der Klis}, M., {M{\'e}ndez}, M., {et~al.} 2008, \apj,
  685, 436

\bibitem[{{Arnaud}(1996)}]{arnaud96}
{Arnaud}, K.~A. 1996, in Astronomical Society of the Pacific Conference Series,
  Vol. 101, Astronomical Data Analysis Software and Systems V, ed. G.~H.
  {Jacoby} \& J.~{Barnes}, 17

\bibitem[{{Ballantyne}(2004)}]{ballan04}
{Ballantyne}, D.~R. 2004, \mnras, 351, 57

\bibitem[{{Ballantyne}(2017)}]{ballantyne17}
{Ballantyne}, D.~R. 2017, \mnras, 472, L60

\bibitem[{{Belloni} {et~al.}(2007){Belloni}, {Homan}, {Motta}, {Ratti}, \&
  {M{\'e}ndez}}]{belloni07}
{Belloni}, T., {Homan}, J., {Motta}, S., {Ratti}, E., \& {M{\'e}ndez}, M. 2007,
  \mnras, 379, 247

\bibitem[{{Beloborodov}(1999)}]{beloborodov99}
{Beloborodov}, A.~M. 1999, \apjl, 510, L123

\bibitem[{{Bhattacharyya} \& {Strohmayer}(2007)}]{Bhattacharyya07}
{Bhattacharyya}, S. \& {Strohmayer}, T.~E. 2007, \apjl, 664, L103

\bibitem[{{Braje} {et~al.}(2000){Braje}, {Romani}, \& {Rauch}}]{braje00}
{Braje}, T.~M., {Romani}, R.~W., \& {Rauch}, K.~P. 2000, \apj, 531, 447

\bibitem[{{Cackett} {et~al.}(2010){Cackett}, {Miller}, {Ballantyne}, {Barret},
  {Bhattacharyya}, {Boutelier}, {Miller}, {Strohmayer}, \&
  {Wijnands}}]{cackett10}
{Cackett}, E.~M., {Miller}, J.~M., {Ballantyne}, D.~R., {et~al.} 2010, \apj,
  720, 205

\bibitem[{{Cackett} {et~al.}(2008){Cackett}, {Miller}, {Bhattacharyya},
  {Grindlay}, {Homan}, {van der Klis}, {Miller}, {Strohmayer}, \&
  {Wijnands}}]{Cackett08}
{Cackett}, E.~M., {Miller}, J.~M., {Bhattacharyya}, S., {et~al.} 2008, \apj,
  674, 415

\bibitem[{{Casares} {et~al.}(2006){Casares}, {Cornelisse}, {Steeghs},
  {Charles}, {Hynes}, {O'Brien}, \& {Strohmayer}}]{casares06}
{Casares}, J., {Cornelisse}, R., {Steeghs}, D., {et~al.} 2006, \mnras, 373,
  1235

\bibitem[{{Chauvin} {et~al.}(2018){Chauvin}, {Flor{\'e}n}, {Friis}, {Jackson},
  {Kamae}, {Kataoka}, {Kawano}, {Kiss}, {Mikhalev}, {Mizuno}, {Ohashi},
  {Stana}, {Tajima}, {Takahashi}, {Uchida}, \& {Pearce}}]{chauvin18}
{Chauvin}, M., {Flor{\'e}n}, H.~G., {Friis}, M., {et~al.} 2018, Nature
  Astronomy, 2, 652

\bibitem[{{Chiang} {et~al.}(2016){Chiang}, {Morgan}, {Cackett}, {Miller},
  {Bhattacharyya}, \& {Strohmayer}}]{chiang16}
{Chiang}, C.-Y., {Morgan}, R.~A., {Cackett}, E.~M., {et~al.} 2016, \apj, 831,
  45

\bibitem[{{D'A{\`i}} {et~al.}(2010){D'A{\`i}}, {di Salvo}, {Ballantyne},
  {Iaria}, {Robba}, {Papitto}, {Riggio}, {Burderi}, {Piraino}, {Santangelo},
  {Matt}, {Dov{\v c}iak}, \& {Karas}}]{daa10}
{D'A{\`i}}, A., {di Salvo}, T., {Ballantyne}, D., {et~al.} 2010, \aap, 516, A36

\bibitem[{{Dauser} {et~al.}(2014){Dauser}, {Garcia}, {Parker}, {Fabian}, \&
  {Wilms}}]{dauser14}
{Dauser}, T., {Garcia}, J., {Parker}, M.~L., {Fabian}, A.~C., \& {Wilms}, J.
  2014, \mnras, 444, L100

\bibitem[{{Dauser} {et~al.}(2016){Dauser}, {Garc{\'{\i}}a}, \&
  {Wilms}}]{dauser16}
{Dauser}, T., {Garc{\'{\i}}a}, J., \& {Wilms}, J. 2016, Astronomische
  Nachrichten, 337, 362

\bibitem[{{Dauser} {et~al.}(2013){Dauser}, {Garcia}, {Wilms}, {B{\"o}ck},
  {Brenneman}, {Falanga}, {Fukumura}, \& {Reynolds}}]{dauser13}
{Dauser}, T., {Garcia}, J., {Wilms}, J., {et~al.} 2013, \mnras, 430, 1694

\bibitem[{{Dauser} {et~al.}(2010){Dauser}, {Wilms}, {Reynolds}, \&
  {Brenneman}}]{dauser10}
{Dauser}, T., {Wilms}, J., {Reynolds}, C.~S., \& {Brenneman}, L.~W. 2010,
  \mnras, 409, 1534

\bibitem[{{Done} {et~al.}(2007){Done}, {Gierli{\'n}ski}, \& {Kubota}}]{done07}
{Done}, C., {Gierli{\'n}ski}, M., \& {Kubota}, A. 2007, \aapr, 15, 1

\bibitem[{{Ferreira} {et~al.}(2006){Ferreira}, {Petrucci}, {Henri},
  {Saug{\'e}}, \& {Pelletier}}]{ferreira06}
{Ferreira}, J., {Petrucci}, P.~O., {Henri}, G., {Saug{\'e}}, L., \&
  {Pelletier}, G. 2006, \aap, 447, 813

\bibitem[{{Ford} {et~al.}(2000){Ford}, {van der Klis}, {M{\'e}ndez},
  {Wijnands}, {Homan}, {Jonker}, \& {van Paradijs}}]{ford2000}
{Ford}, E.~C., {van der Klis}, M., {M{\'e}ndez}, M., {et~al.} 2000, \apj, 537,
  368

\bibitem[{{Fukumura} \& {Kazanas}(2007)}]{Fukumura07}
{Fukumura}, K. \& {Kazanas}, D. 2007, \apj, 664, 14

\bibitem[{{Galloway} {et~al.}(2006){Galloway}, {Psaltis}, {Muno}, \&
  {Chakrabarty}}]{galloway06}
{Galloway}, D.~K., {Psaltis}, D., {Muno}, M.~P., \& {Chakrabarty}, D. 2006,
  \apj, 639, 1033

\bibitem[{{Garc{\'{\i}}a} {et~al.}(2014){Garc{\'{\i}}a}, {Dauser}, {Lohfink},
  {Kallman}, {Steiner}, {McClintock}, {Brenneman}, {Wilms}, {Eikmann},
  {Reynolds}, \& {Tombesi}}]{garcia14}
{Garc{\'{\i}}a}, J., {Dauser}, T., {Lohfink}, A., {et~al.} 2014, \apj, 782, 76

\bibitem[{{Garc{\'{\i}}a} {et~al.}(2013){Garc{\'{\i}}a}, {Dauser}, {Reynolds},
  {Kallman}, {McClintock}, {Wilms}, \& {Eikmann}}]{garcia13}
{Garc{\'{\i}}a}, J., {Dauser}, T., {Reynolds}, C.~S., {et~al.} 2013, \apj, 768,
  146

\bibitem[{{Garc{\'\i}a} \& {Kallman}(2010)}]{garcia10}
{Garc{\'\i}a}, J. \& {Kallman}, T.~R. 2010, \apj, 718, 695

\bibitem[{{Garc{\'\i}a} {et~al.}(2022){Garc{\'\i}a}, {Dauser}, {Ludlam},
  {Parker}, {Fabian}, {Harrison}, \& {Wilms}}]{garcia22}
{Garc{\'\i}a}, J.~A., {Dauser}, T., {Ludlam}, R., {et~al.} 2022, \apj, 926, 13

\bibitem[{{Ghisellini} {et~al.}(1991){Ghisellini}, {George}, {Fabian}, \&
  {Done}}]{Ghisell91}
{Ghisellini}, G., {George}, I.~M., {Fabian}, A.~C., \& {Done}, C. 1991, \mnras,
  248, 14

\bibitem[{{Giles} {et~al.}(2002){Giles}, {Hill}, {Strohmayer}, \&
  {Cummings}}]{giles02}
{Giles}, A.~B., {Hill}, K.~M., {Strohmayer}, T.~E., \& {Cummings}, N. 2002,
  \apj, 568, 279

\bibitem[{{Haardt}(1993)}]{haardt93}
{Haardt}, F. 1993, \apj, 413, 680

\bibitem[{{Haardt} \& {Maraschi}(1991)}]{haardt91}
{Haardt}, F. \& {Maraschi}, L. 1991, \apjl, 380, L51

\bibitem[{{Haardt} {et~al.}(1994){Haardt}, {Maraschi}, \&
  {Ghisellini}}]{haardt94}
{Haardt}, F., {Maraschi}, L., \& {Ghisellini}, G. 1994, \apjl, 432, L95

\bibitem[{{Hasinger} \& {van der Klis}(1989)}]{hasinger89}
{Hasinger}, G. \& {van der Klis}, M. 1989, \aap, 225, 79

\bibitem[{{Hiemstra} {et~al.}(2011){Hiemstra}, {M{\'e}ndez}, {Done},
  {D{\'{\i}}az Trigo}, {Altamirano}, \& {Casella}}]{hiemstra11}
{Hiemstra}, B., {M{\'e}ndez}, M., {Done}, C., {et~al.} 2011, \mnras, 411, 137

\bibitem[{{Homan} {et~al.}(2007){Homan}, {van der Klis}, {Wijnands}, {Belloni},
  {Fender}, {Klein-Wolt}, {Casella}, {M{\'e}ndez}, {Gallo}, {Lewin}, \&
  {Gehrels}}]{homan07}
{Homan}, J., {van der Klis}, M., {Wijnands}, R., {et~al.} 2007, \apj, 656, 420

\bibitem[{{Jahoda} {et~al.}(2006){Jahoda}, {Markwardt}, {Radeva}, {Rots},
  {Stark}, {Swank}, {Strohmayer}, \& {Zhang}}]{jahoda06}
{Jahoda}, K., {Markwardt}, C.~B., {Radeva}, Y., {et~al.} 2006, \apjs, 163, 401

\bibitem[{{Kaaret} {et~al.}(1999){Kaaret}, {Piraino}, {Ford}, \&
  {Santangelo}}]{kaaret99}
{Kaaret}, P., {Piraino}, S., {Ford}, E.~C., \& {Santangelo}, A. 1999, \apjl,
  514, L31

\bibitem[{{Kinch} {et~al.}(2016){Kinch}, {Schnittman}, {Kallman}, \&
  {Krolik}}]{kinch16}
{Kinch}, B.~E., {Schnittman}, J.~D., {Kallman}, T.~R., \& {Krolik}, J.~H. 2016,
  \apj, 826, 52

\bibitem[{{Kinch} {et~al.}(2019){Kinch}, {Schnittman}, {Kallman}, \&
  {Krolik}}]{kinch19}
{Kinch}, B.~E., {Schnittman}, J.~D., {Kallman}, T.~R., \& {Krolik}, J.~H. 2019,
  \apj, 873, 71

\bibitem[{{Krolik}(1999)}]{krolik99}
{Krolik}, J.~H. 1999, {Active galactic nuclei : from the central black hole to
  the galactic environment}

\bibitem[{{Ludlam} {et~al.}(2020){Ludlam}, {Cackett}, {Garc{\'\i}a}, {Miller},
  {Bult}, {Strohmayer}, {Guillot}, {Jaisawal}, {Malacaria}, {Fabian}, \&
  {Markwardt}}]{ludlam20}
{Ludlam}, R.~M., {Cackett}, E.~M., {Garc{\'\i}a}, J.~A., {et~al.} 2020, \apj,
  895, 45

\bibitem[{{Ludlam} {et~al.}(2022){Ludlam}, {Cackett}, {Garc{\'\i}a}, {Miller},
  {Stevens}, {Fabian}, {Homan}, {Ng}, {Guillot}, {Buisson}, \&
  {Chakrabarty}}]{ludlam22}
{Ludlam}, R.~M., {Cackett}, E.~M., {Garc{\'\i}a}, J.~A., {et~al.} 2022, \apj,
  927, 112

\bibitem[{{Ludlam} {et~al.}(2018){Ludlam}, {Miller}, {Arzoumanian}, {Bult},
  {Cackett}, {Chakrabarty}, {Dauser}, {Enoto}, {Fabian}, {Garc{\'\i}a},
  {Gendreau}, {Guillot}, {Homan}, {Jaisawal}, {Keek}, {La Marr}, {Malacaria},
  {Markwardt}, {Steiner}, \& {Strohmayer}}]{ludlam18}
{Ludlam}, R.~M., {Miller}, J.~M., {Arzoumanian}, Z., {et~al.} 2018, \apjl, 858,
  L5

\bibitem[{{Ludlam} {et~al.}(2017){Ludlam}, {Miller}, {Bachetti}, {Barret},
  {Bostrom}, {Cackett}, {Degenaar}, {Di Salvo}, {Natalucci}, {Tomsick},
  {Paerels}, \& {Parker}}]{ludlam17}
{Ludlam}, R.~M., {Miller}, J.~M., {Bachetti}, M., {et~al.} 2017, \apj, 836, 140

\bibitem[{{Lyu} {et~al.}(2014){Lyu}, {M{\'e}ndez}, {Sanna}, {Homan}, {Belloni},
  \& {Hiemstra}}]{lyu14}
{Lyu}, M., {M{\'e}ndez}, M., {Sanna}, A., {et~al.} 2014, \mnras, 440, 1165

\bibitem[{{Lyu} {et~al.}(2019){Lyu}, {M{\'e}ndez}, {Zhang}, \& {Xiang}}]{lyu19}
{Lyu}, M., {M{\'e}ndez}, M., {Zhang}, J.-F., \& {Xiang}, F.-Y. 2019, \mnras,
  484, 3434

\bibitem[{{Makishima} {et~al.}(1986){Makishima}, {Maejima}, {Mitsuda}, {Bradt},
  {Remillard}, {Tuohy}, {Hoshi}, \& {Nakagawa}}]{maki86}
{Makishima}, K., {Maejima}, Y., {Mitsuda}, K., {et~al.} 1986, \apj, 308, 635

\bibitem[{{Malu} {et~al.}(2020){Malu}, {Sriram}, \& {Agrawal}}]{malu20}
{Malu}, S., {Sriram}, K., \& {Agrawal}, V.~K. 2020, \mnras, 499, 2214

\bibitem[{{Markoff} {et~al.}(2001){Markoff}, {Falcke}, \& {Fender}}]{markoff01}
{Markoff}, S., {Falcke}, H., \& {Fender}, R. 2001, \aap, 372, L25

\bibitem[{{Markoff} {et~al.}(2005){Markoff}, {Nowak}, \& {Wilms}}]{markoff05}
{Markoff}, S., {Nowak}, M.~A., \& {Wilms}, J. 2005, \apj, 635, 1203

\bibitem[{{M{\'e}ndez} \& {van der Klis}(1999)}]{mendez99}
{M{\'e}ndez}, M. \& {van der Klis}, M. 1999, \apjl, 517, L51

\bibitem[{{Mitsuda} {et~al.}(1984){Mitsuda}, {Inoue}, {Koyama}, {Makishima},
  {Matsuoka}, {Ogawara}, {Suzuki}, {Tanaka}, {Shibazaki}, \&
  {Hirano}}]{mitsuda84}
{Mitsuda}, K., {Inoue}, H., {Koyama}, K., {et~al.} 1984, \pasj, 36, 741

\bibitem[{{Mondal} {et~al.}(2021){Mondal}, {Raychaudhuri}, \&
  {Dewangan}}]{mondal21}
{Mondal}, A.~S., {Raychaudhuri}, B., \& {Dewangan}, G.~C. 2021, \mnras, 504,
  1331

\bibitem[{{Narayan} \& {McClintock}(2008)}]{narayan08}
{Narayan}, R. \& {McClintock}, J.~E. 2008, \nar, 51, 733

\bibitem[{{Ng} {et~al.}(2010){Ng}, {D{\'{\i}}az Trigo}, {Cadolle Bel}, \&
  {Migliari}}]{ng10}
{Ng}, C., {D{\'{\i}}az Trigo}, M., {Cadolle Bel}, M., \& {Migliari}, S. 2010,
  \aap, 522, A96

\bibitem[{{Pandel} {et~al.}(2008){Pandel}, {Kaaret}, \& {Corbel}}]{pandel08}
{Pandel}, D., {Kaaret}, P., \& {Corbel}, S. 2008, \apj, 688, 1288

\bibitem[{{Penna} {et~al.}(2013){Penna}, {S{\c a}dowski}, {Kulkarni}, \&
  {Narayan}}]{penna13}
{Penna}, R.~F., {S{\c a}dowski}, A., {Kulkarni}, A.~K., \& {Narayan}, R. 2013,
  \mnras, 428, 2255

\bibitem[{{Rothschild} {et~al.}(1998){Rothschild}, {Blanco}, {Gruber},
  {Heindl}, {MacDonald}, {Marsden}, {Pelling}, {Wayne}, \& {Hink}}]{roth98}
{Rothschild}, R.~E., {Blanco}, P.~R., {Gruber}, D.~E., {et~al.} 1998, \apj,
  496, 538

\bibitem[{{Saavedra} {et~al.}(2023){Saavedra}, {Garc{\'\i}a}, {Fogantini},
  {M{\'e}ndez}, {Combi}, {Luque-Escamilla}, \& {Mart{\'\i}}}]{saavedra2023}
{Saavedra}, E.~A., {Garc{\'\i}a}, F., {Fogantini}, F.~A., {et~al.} 2023,
  \mnras, 522, 3367

\bibitem[{{Sanna} {et~al.}(2013){Sanna}, {Hiemstra}, {M{\'e}ndez},
  {Altamirano}, {Belloni}, \& {Linares}}]{sanna13}
{Sanna}, A., {Hiemstra}, B., {M{\'e}ndez}, M., {et~al.} 2013, \mnras, 432, 1144

\bibitem[{{Shih} {et~al.}(2005){Shih}, {Bird}, {Charles}, {Cornelisse}, \&
  {Tiramani}}]{shih05}
{Shih}, I.~C., {Bird}, A.~J., {Charles}, P.~A., {Cornelisse}, R., \&
  {Tiramani}, D. 2005, \mnras, 361, 602

\bibitem[{{Strohmayer} \& {Markwardt}(2002)}]{strohmayer02}
{Strohmayer}, T.~E. \& {Markwardt}, C.~B. 2002, \apj, 577, 337

\bibitem[{{Str{\"u}der} {et~al.}(2001){Str{\"u}der}, {Briel}, {Dennerl},
  {Hartmann}, {Kendziorra}, {Meidinger}, {Pfeffermann}, {Reppin}, {Aschenbach},
  {Bornemann}, {Br{\"a}uninger}, {Burkert}, {Elender}, {Freyberg}, {Haberl},
  {Hartner}, {Heuschmann}, {Hippmann}, {Kastelic}, {Kemmer}, {Kettenring},
  {Kink}, {Krause}, {M{\"u}ller}, {Oppitz}, {Pietsch}, {Popp}, {Predehl}, \&
  {Read}}]{struder01}
{Str{\"u}der}, L., {Briel}, U., {Dennerl}, K., {et~al.} 2001, \aap, 365, L18

\bibitem[{{van Paradijs} {et~al.}(1990){van Paradijs}, {van der Klis}, {van
  Amerongen}, {Pedersen}, {Smale}, {Mukai}, {Schoembs}, {Haefner}, {Pfeiffer},
  \& {Lewin}}]{van90}
{van Paradijs}, J., {van der Klis}, M., {van Amerongen}, S., {et~al.} 1990,
  \aap, 234, 181

\bibitem[{{Vasudevan} \& {Fabian}(2007)}]{vasud07}
{Vasudevan}, R.~V. \& {Fabian}, A.~C. 2007, \mnras, 381, 1235

\bibitem[{{Verner} {et~al.}(1996){Verner}, {Ferland}, {Korista}, \&
  {Yakovlev}}]{verner96}
{Verner}, D.~A., {Ferland}, G.~J., {Korista}, K.~T., \& {Yakovlev}, D.~G. 1996,
  \apj, 465, 487

\bibitem[{{Wang} {et~al.}(2021){Wang}, {Mastroserio}, {Kara}, {Garc{\'\i}a},
  {Ingram}, {Connors}, {van der Klis}, {Dauser}, {Steiner}, {Buisson}, {Homan},
  {Lucchini}, {Fabian}, {Bright}, {Fender}, {Cackett}, \& {Remillard}}]{wang21}
{Wang}, J., {Mastroserio}, G., {Kara}, E., {et~al.} 2021, \apjl, 910, L3

\bibitem[{{Wang} {et~al.}(2017){Wang}, {M{\'e}ndez}, {Sanna}, {Altamirano}, \&
  {Belloni}}]{wyn17}
{Wang}, Y., {M{\'e}ndez}, M., {Sanna}, A., {Altamirano}, D., \& {Belloni},
  T.~M. 2017, \mnras, 468, 2256

\bibitem[{{Wilkins}(2018)}]{wilkin18}
{Wilkins}, D.~R. 2018, \mnras, 475, 748

\bibitem[{{Wilkins} \& {Fabian}(2011)}]{wilkins11}
{Wilkins}, D.~R. \& {Fabian}, A.~C. 2011, \mnras, 414, 1269

\bibitem[{{Wilms} {et~al.}(2000){Wilms}, {Allen}, \& {McCray}}]{wilms00}
{Wilms}, J., {Allen}, A., \& {McCray}, R. 2000, \apj, 542, 914

\bibitem[{{Zdziarski} \& {Gierli{\'n}ski}(2004)}]{zdziarski04}
{Zdziarski}, A.~A. \& {Gierli{\'n}ski}, M. 2004, Progress of Theoretical
  Physics Supplement, 155, 99

\bibitem[{{Zhang} {et~al.}(2011){Zhang}, {M{\'e}ndez}, \&
  {Altamirano}}]{zhang11}
{Zhang}, G., {M{\'e}ndez}, M., \& {Altamirano}, D. 2011, \mnras, 413, 1913

\bibitem[{{Zhang} {et~al.}(2009){Zhang}, {M{\'e}ndez}, {Altamirano}, {Belloni},
  \& {Homan}}]{zhang09}
{Zhang}, G., {M{\'e}ndez}, M., {Altamirano}, D., {Belloni}, T.~M., \& {Homan},
  J. 2009, \mnras, 398, 368

\bibitem[{{Zhang} {et~al.}(2019){Zhang}, {Dov{\v{c}}iak}, \&
  {Bursa}}]{zhangwd19}
{Zhang}, W., {Dov{\v{c}}iak}, M., \& {Bursa}, M. 2019, \apj, 875, 148

\bibitem[{{Zhang} {et~al.}(1997){Zhang}, {Lapidus}, {Swank}, {White}, \&
  {Titarchuk}}]{zhang97}
{Zhang}, W., {Lapidus}, I., {Swank}, J.~H., {White}, N.~E., \& {Titarchuk}, L.
  1997, \iaucirc, 6541, 1

\bibitem[{{Zhang} {et~al.}(1996){Zhang}, {Lapidus}, {White}, \&
  {Titarchuk}}]{zhang96}
{Zhang}, W., {Lapidus}, I., {White}, N.~E., \& {Titarchuk}, L. 1996, \apjl,
  469, L17

\end{thebibliography}
%\end{thebibliograghy}
%

%\clearpage
\begin{appendix}
\section{Comparison with the results from the $kT_{\rm BB}$-corrected model NTHRATIO}

Based on the  {\sc xillver} models, the reflection model {\sc relxill} also fixed the temperature of the seed photons for the thermal Comptonization at 0.05 keV or 0.01 keV, which are appropriate for black hole systems, but much lower than the expected value in a neutron star system. In order to see the possible influence induced by the low temperature, we further fitted the spectra with the model {\sc bbody+diskbb+nthcomp+relconv$\times$(xillvercp$\times$nthratio)}, where the model {\sc nthratio}\footnote{https://github.com/garciafederico/nthratio} is designed to make a first-order correction to the soft-excess introduced by the  {\tt xillver}-based models. To compute this correction, {\sc nthratio} uses as inputs the power-law index, $\Gamma$; the electron temperature, kT$_{e}$; and the seed photon temperature, $kT_{\rm BB}$, from the {\sc nthcomp} model, and thus no new free parameters need to be added to the fits. The reflection fraction parameter in {\sc xillverCp} is set to -1 so that only the reflection is returned. 

As shown in Table \ref{nthratio}, we found that the fitted parameters in the two hard observations (X1 and X6) are approximately the same as in the   {\sc relxillCp} model (Table \ref{relxill}). For the soft observations (X2--X4), the most significant difference comes from the ionization parameter, log($\xi$), which yields higher values than in {\sc relxillCp}. Additionally, the hydrogen column, $N_{\rm H}$, is lower compared with those in {\sc relxillCp}, and the disk temperature, $kT_{\rm disk}$, is a little lower than the values in the  {\sc relxillCp} model. Those parameter differences do not make changes to the conclusions in this work. The main parameter that likely influences the conclusions, log($\xi$), is much larger in the model {\sc nthratio} in the soft state, leading to an even bigger gap between the fitted values and the predicted ionization curves from the theory.

%\begin{landscape}
\begin{table*}
\tiny
\centering
\caption{Best-fitting results for the fit to the X-ray spectra of 4U 1636--53 with the reflection model {\sc bbody+diskbb+nthcomp+relconv$\times$(xillvercp$\times$nthratio)}.}

\begin{tabular}{cccccccc}
\hline
\hline
Model Comp   &   Parameter     & X1       & X2       & X3       & X4       & X5        & X6    \\
\hline

{\sc Tbabs}    &$N_{\rm H}$ (10$^{22}$cm$^{-2}$)      &   0.52$\pm$0.06              &    0.23$\pm$0.05            &       0.28$\pm$0.04               &    0.31$\pm$0.03            &   0.29$\pm$0.03                    &   0.49$_{-0.08}^{+0.05}$                                          \\                                                                                                         
\\
{\sc bbody}    &$kT_{\rm BB}$ (keV)                   &   1.84$_{-0.11}^{+0.21}$     &    2.14$_{-0.08}^{+0.14}$   &       2.12$\pm$0.03               &    2.11$\pm$0.04            &   2.15$_{-0.06}^{+0.03}$           &   2.39$\pm$0.08                                                   \\
               &Norm       (10$^{-3}$)                &   1.0$\pm$0.1                &    6.7$\pm$0.7              &       9.3$_{-0.3}^{+0.9}$         &    8.9$\pm$0.4              &   10.6$\pm$1.1                     &   3.1$\pm$0.3                                                   \\                                                                                                          
               &Flux (10$^{-10}$ c.g.s)               &   0.92$_{  -0.13  }^{+  0.16   }$    &   5.09$_{  -0.80  }^{+  1.14   }$    &    7.72$_{  -0.29  }^{+  1.07   }$     &   7.31$_{  -0.21  }^{+  0.27   }$    &    9.34$_{  -1.29  }^{+  0.78   }$    &  2.76$_{  -0.40  }^{+  0.12   }$     \\   
\\          
{\sc diskbb}   &$kT_{\rm disk}$ (keV)                 &   0.20$\pm$0.01              &    0.73$_{-0.11}^{+0.05}$   &       0.78$\pm$0.06               &    0.70$_{-0.02}^{+0.10}$   &   0.86$\pm$0.06                    &   0.23$\pm$0.02                                                   \\
               &Norm                                  &   15949$_{-7904}^{+18732}$   &    85$_{-36}^{+31}$         &       <20                         &    <52                      &   <60                              &   8072$_{-4855}^{+6526}$                                          \\                                                                                                           
               &Flux (10$^{-10}$ c.g.s)               &4.60$_{  -3.44  }^{+  4.84   }$    &   5.37$_{  -3.42  }^{+  2.97   }$    &      <2.14      &   0.28$_{  -0.28  }^{+  8.60   }$    &    1.90$_{  -1.90  }^{+  2.75   }$    &  2.53$_{  -1.90  }^{+  2.19   }$   \\     
\\
{\sc nthcomp}  &$\Gamma$                              &   1.83$\pm$0.01              &    2.23$_{-0.10}^{+0.35}$   &       2.59$\pm$0.13               &    2.93$_{-0.61}^{+0.07}$   &   2.58$_{-0.12}^{+0.41}$           &   1.89$\pm$0.02                                                \\
               &kT$_{e}$ (keV)                        &   17$_{-2}^{+15}$            &    8$_{-2}^{+12}$           &       6$_{-1}^{+2}$               &    15$_{-10}^{+35}$         &   5$_{-1}^{+5}$                    &   44$_{-12}^{+36*}$                                                    \\
               &Norm                                  &   0.10$_{-0.07}^{+0.03}$     &    <0.16                    &       0.15$_{-0.09}^{+0.05}$      &    0.23$_{-0.11}^{+0.02}$   &   0.0376$_{-0.0375*}^{+0.2182}$    &   0.02$_{-0.01}^{+0.04}$                                          \\
               &Flux (10$^{-10}$ c.g.s)               &13.90$_{   -7.17 }^{+   2.75  }$   &   0.20$_{  -0.20  }^{+  10.49  }$    &    10.03$_{   -8.07 }^{+   5.15  }$    &   12.98$_{   -8.40 }^{+   1.70  }$   &    1.68$_{  -1.68  }^{+  12.75  }$    &  2.39$_{  -2.39  }^{+  4.96   }$      \\  
\\
{\sc relconv}  &$\beta$                               &   1.8$\pm$0.8                &    2.1$_{-0.3}^{+0.2}$      &       2.4$\pm$0.2                 &    2.3$\pm$0.1              &   2.4$\pm$0.1                      &   1.8$_{-0.3}^{+0.4}$                                          \\   
               &$R_{\rm in}$ ($R_{\rm g}$)            &   9.22$_{-4.10}^{+40.78}$    &    5.12$_{-0*}^{+1.42}$     &       7.10$_{-1.74}^{+3.15}$      &    5.12$_{-0*}^{+0.44}$     &   5.12$_{-0*}^{+0.39}$             &   5.14$_{-0.02*}^{+5.43}$                                     \\
\\
{\sc xillverCp}&log($\xi$)                            &   3.61$_{-0.13}^{+0.10}$     &    4.10$_{-0.28}^{+0.41}$   &       4.17$_{-0.19}^{+0.11}$      &    4.01$_{-0.30}^{+0.25}$   &   4.15$_{-0.38}^{+0.18}$           &   3.66$_{-0.04}^{+0.02}$                                        \\
               &Norm (10$^{-3}$)                      &   3.0$_{-0.8}^{+1.4}$        &    10.5$_{-3.0}^{+4.6}$     &       27.9$_{-4.7}^{+6.1}$        &    21.7$_{-5.3}^{+7.0}$     &   37.3$_{-18.7}^{+9.1}$            &   7.7$_{-0.6}^{+1.0}$                                             \\
               &Flux (10$^{-10}$ c.g.s)               &10.04$_{   -3.54 }^{+   3.96  }$   &   15.65$_{   -7.95 }^{+   10.87 }$   &    19.43$_{   -5.40 }^{+   6.16  }$    &   8.75$_{  -1.92  }^{+  13.18  }$    &    21.72$_{   -8.82 }^{+   3.08  }$   &  26.67$_{   -4.60 }^{+   4.04  }$   \\                                  
\hline
               &$\chi^2_\nu$($\chi^2/dof)$     & 0.97 (172/177)  & 0.97 (152/157)  & 1.11 (173/156)   & 0.86 (135/157)  & 1.08 (167/155)  & 1.21 (216/179)  \\
               &Total flux (10$^{-10}$ c.g.s)  & 29.99$_{-5.26}^{+3.66}$            &   28.83$_{   -0.76 }^{+   0.36  }$   &    36.54$_{   -0.62 }^{+   0.35  }$    &   28.72$_{   -0.45 }^{+   0.43  }$   &    33.74$_{   -0.41 }^{+   0.35  }$   &  34.30$_{   -2.71 }^{+   1.71  }$       \\

\hline
\end{tabular} 
\medskip  
\label{nthratio}
\end{table*}
%\end{landscape}

\begin{table*}
\centering
\caption{Best-fitting results for the fit to the X-ray spectra of 4U 1636--53 with the reflection model {\sc bbody+diskbb+relxilllp}. In the fits we fixed the reflection fraction parameter at 1 and 2 for the observations in the hard and soft spectra state, respectively.}
\begin{tabular}{cccccccc}
\hline
\hline
Model Comp   &   Parameter     & X1       & X2       & X3       & X4       & X5        & X6    \\
\hline

{\sc Tbabs}    &$N_{\rm H}$ (10$^{22}$cm$^{-2}$)      &    0.47$\pm$0.04            & 0.39$\pm$0.04             & 0.46$\pm$0.03             & 0.43$\pm$0.05            &  0.44$\pm$0.03               &  0.42$_{-0.08}^{+0.04}$                                                  \\
\\
{\sc bbody}    &$kT_{\rm BB}$ (keV)                   &    1.69$\pm$0.13            & 2.06$\pm$0.05             & 2.00$\pm$0.02             & 2.01$\pm$0.03            &  2.03$\pm$0.02               &  2.23$_{-0.17}^{+ 0.14}$                                                  \\
               &Norm       (10$^{-3}$)                &    0.61$\pm$0.17            & 6.86$_{-0.35}^{+0.28}$    & 12.01$_{-0.41}^{+0.25}$   & 10.91$\pm$0.37           &  12.44$_{-0.14}^{+0.21}$     &  1.59$_{-0.38}^{+ 0.22}$                                                  \\                                                                                                                                                                         
\\
{\sc diskbb}   &$kT_{\rm disk}$ (keV)                 &    0.23$\pm$0.02            & 0.83$\pm$0.02             & 0.94$\pm$0.02             & 0.91$\pm$0.03            &  0.92$\pm$0.01               &  0.32$_{-0.03}^{+ 0.07}$                                                  \\
               &Norm                                  &    5690$_{-2712}^{+4183}$   & 81$\pm$12                 & 75$_{-6}^{+9}$            & 72$\pm$7                 &  92$_{ -4}^{+7}$             &  1562$_{-957}^{+1341}$                                                      \\                                                                                                                                                                                
\\
{\sc relxilllp}&$h$ ($R_{\rm g}$)                     &    42$_{-19}^{+28}$         & 30$_{-19}^{+21}$          & 29$_{-24}^{+10}$          & 25$_{-9}^{+23}$          &  23$_{-5}^{+8}$              &   43$_{-17}^{+21}$                                                       \\
               &$R_{\rm in}$ ($R_{\rm g}$)            &    5.12$_{-0*}^{+19.59}$    & 5.12$_{-0*}^{+17.41}$     & 5.80$_{-0.68*}^{+17.69}$   & 5.12$_{-0*}^{+2.73}$     &  5.12$_{-0*}^{+1.51}$        &  5.12$_{-0*}^{+15.78}$                                                           \\
               &$\Gamma$                              &    1.69$\pm$0.03            & 1.84$\pm$0.08             & 2.00$_{-0.08}^{+0.04}$    & 2.01$_{-0.44}^{+0.08}$   &  1.99$_{-0.08}^{+0.04}$      &  1.67$\pm$0.07                                                         \\
               &log($\xi$)                            &    3.42$\pm$0.05            & 3.64$_{-0.12}^{+0.08}$    & 3.46$\pm$0.17             & 3.43$_{-0.20}^{+0.24}$   &  3.30$\pm$0.05               &  3.31$\pm$0.05                                                            \\
               &E$_{cut}$ (keV)                       &    53$_{-8}^{+11}$          & 19$\pm$3                  & 19$_{-3}^{+1}$            & 20$_{-3}^{+6}$           &  19$\pm$3                    &  41$_{-8}^{+13}$                                                       \\
               &R$_{refl}$                            &      1      &   2      &  2   &  2    &     2    &   1                                              \\

               &Norm (10$^{-4}$)                      &    22.02$_{-0.76}^{+3.16}$  & 13.98$_{-2.09}^{+2.53}$   & 17.35$_{-1.92}^{+15.94}$  & 11.82$_{-3.76}^{+2.43}$  &  13.02$_{-2.52}^{+1.29}$     &  26.39$_{-0.98}^{+2.38}$                                                  \\

\hline
\end{tabular}

\medskip  
\label{fix}
\end{table*}

\end{appendix}

\end{document}